\documentclass[12pt]{article}
\usepackage{amsfonts,amsmath,amssymb,mathrsfs}

\title{Bell-Type Quantum Field Theories}
\author{
   Detlef D\"urr\footnote{Mathematisches Institut der Universit\"{a}t
         M\"{u}nchen, Theresienstra{\ss}e 39, 80333 M\"{u}nchen, Germany.
         E-mail: duerr@mathematik.uni-muenchen.de},
   Sheldon Goldstein\footnote{Departments of Mathematics and Physics -
         Hill Center, Rutgers, The State University of New Jersey,
         110 Frelinghuysen Road, Piscataway, NJ 08854-8019, USA.
         E-mail: oldstein@math.rutgers.edu},\\
   Roderich Tumulka\footnote{Dipartimento di Fisica and INFN sezione di
         Genova,
         Via Dodecaneso 33, 16146 Genova, Italy.
         E-mail: tumulka@mathematik.uni-muenchen.de}, and
   Nino Zangh\`\i\footnote{Dipartimento di Fisica and INFN sezione di
         Genova,
         Via Dodecaneso 33, 16146 Genova, Italy. E-mail: 
         zanghi@ge.infn.it}
}
\date{July 15, 2004}

\newcommand{\CCC}{\mathbb{C}} 
\newcommand{\RRR}{\mathbb{R}} 
\newcommand{\NNN}{\mathbb{N}} 
\newcommand{\ZZZ}{\mathbb{Z}} 
\newcommand{\EEE}{\mathbb{E}} 
\newcommand{\E}{e} 
\newcommand{\I}{i} 
\newcommand{\1}{\mathbf{1}} 
\newcommand{\tr}{\mathrm{tr}} 
\newcommand{\Laplace}{\Delta} 
\newcommand{\ov}{\overline}
\renewcommand{\Re}{\mathrm{Re}} 
\renewcommand{\Im}{\mathrm{Im}} 
\newcommand{\Anti}{{\mathrm{Anti}\,}} 

\newcommand{\Hilbert}{\mathscr{H}}
\renewcommand{\sp}[2]{\langle #1 | #2 \rangle} 
\newcommand{\scalar}[2]{\langle\!\langle #1 | #2 \rangle\!\rangle} %
\newcommand{\Fock}{\mathscr{F}} 
\newcommand{\conf}{\mathcal{Q}} 
\renewcommand{\div}{\,\mathrm{div}\,} 
\newcommand{\prob}{\mathrm{Prob}}
\newcommand{\measure}{\mathbb{P}} 
\newcommand{\current}{\mathbb{J}}
\newcommand{\generator}{\mathscr{L}} 
\newcommand{\pov}{{P}}
\newcommand{\profile}{\varphi}

\newcommand{\covering}{\pi} 
\newcommand{\permutation}{\varrho} 
\newcommand{\Gommo}{\Gamma_{\!\neq}} 

\newcommand{\Dindex}{i} 
\newcommand{\Dindextwo}{j} 
\newcommand{\Dindexthree}{k}
\newcommand{\Dindexfour}{\ell}
\newcommand{\pDindex}{{\tilde{\imath}}}

\newcommand{\ve}{{\boldsymbol e}}

\newcommand{\vX}{\boldsymbol X}
\newcommand{\vx}{{\boldsymbol x}} 
\newcommand{\vy}{{\boldsymbol y}}

\newcommand{\vq}{{\boldsymbol q}}
\newcommand{\vQ}{{\boldsymbol Q}}
\newcommand{\vA}{{\boldsymbol A}}
\newcommand{\valpha}{{\boldsymbol \alpha}}

\newcommand{\vr}{{\boldsymbol r}}

\newcommand{\pvx}{{\widetilde{\vx}}}
\newcommand{\px}{{\widetilde{x}}}
\newcommand{\pn}{{\widetilde{n}}}
\newcommand{\pN}{{\widetilde{N}}}
\newcommand{\ph}{\widetilde{h}}
\newcommand{\pk}{{\widetilde{k}}}
\newcommand{\pvX}{\widetilde{\boldsymbol X}}
\newcommand{\palpha}{\widetilde\valpha}
\newcommand{\ppermutation}{{\widetilde{\permutation}}}

\newcommand{\inter}{{I}} 
\newcommand{\el}{\mathrm{e}} 
\newcommand{\pos}{\mathrm{p}} 
\newcommand{\cond}{\mathrm{cond}} 

\newcommand{\ext}{{\mathrm{ext}}} 
\newcommand{\crea}{{\mathrm{crea}}} 
\newcommand{\ann}{{\mathrm{ann}}} 
\newcommand{\fer}{{\mathrm{f}}} 
\newcommand{\bos}{{\mathrm{b}}} 

\begin{document}\maketitle
\begin{abstract}
In \cite{BellBeables} John~S.~Bell proposed how to associate particle
trajectories with a lattice quantum field theory, yielding what can be
regarded as a $|\Psi|^2$-distributed Markov process on the appropriate
configuration space. A similar process can be defined in the
continuum, for more or less any regularized quantum field theory; such
processes we call Bell-type quantum field theories. We describe
methods for explicitly constructing these processes.  These concern,
in addition to the definition of the Markov processes, the efficient
calculation of jump rates, how to obtain the process from the
processes corresponding to the free and interaction Hamiltonian alone,
and how to obtain the free process from the free Hamiltonian or,
alternatively, from the one-particle process by a construction
analogous to ``second quantization.''  As an example, we consider the
process for a second quantized Dirac field in an external
electromagnetic field.

\medskip

\end{abstract}

\tableofcontents

\section{Introduction}

The aim of this paper is to present methods for constructing Bell-type
QFTs.  The primary variables of Bell-type QFTs are the positions of
the particles.  Bell suggested a dynamical law, governing the motion
of the particles, in which the Hamiltonian $H$ and the state vector
$\Psi$ determine certain jump rates \cite{BellBeables}.  Since these
rates are in a sense the smallest choice possible, we call them the
\emph{minimal jump rates}.  By construction, they preserve the
$|\Psi|^2$ distribution.  We assume a well-defined Hamiltonian as
given; to achieve this, it is often necessary to introduce
cut-offs. We shall assume this has been done where needed.  In cases
in which one has to choose between several possible position
observables, for example because of issues related to the
Newton--Wigner operator \cite{NewtonWigner,Haag}, we shall also assume
that a choice has been made.

Bell-type QFTs can also be regarded as extensions of Bohmian
mechanics.  When one tries to incorporate particle creation and
annihilation into Bohmian mechanics, one is naturally lead to models
like the one we presented in \cite{crea1}.  The quantum equilibrium
distribution, playing a central role in Bohmian mechanics, then more
or less dictates that creation of a particle occurs in a stochastic
manner---just as in Bell's model.

Bell-type QFTs have in common a good deal of mathematical structure,
which we will elucidate.  The paper is organized as follows.  In
Section 2 we introduce all the main ideas and reasonings; a
superficial reading should focus on this section.  Some examples of
Bell-type QFTs are presented in Section 3. (Simple examples of minimal
jump rates can be found in \cite{crea2A}.)  In Section 4 we describe
the construction of a process for the free Hamiltonian based on
``second quantization.'' In Section 5 we sketch the concept of the
``minimal process'' associated with a Hamiltonian $H$. Section 6
concerns some properties of Bell-type QFTs that derive from the
construction methods developed in this paper.  In Section 7 we
conclude.

\section{Ingredients of Bell-Type Quantum Field Theories}
\label{sec:making}

\subsection{Review of Bohmian Mechanics and Equivariance}

Bohmian mechanics \cite{Bohm52,DGZ,Stanford} is a non-relativistic
theory about $N$ point particles moving in 3-space, according to which
the configuration $Q=(\vQ_1,\ldots,\vQ_N)$ evolves according
to\footnote{ The masses $m_k$ of the particles have been absorbed in
the Riemann metric $g_{\mu\nu}$ on configuration space $\RRR^{3N}$,
$g_{ia,jb} = m_i \, \delta_{ij}\, \delta_{ab}$, $i,j=1\ldots N, \:
a,b=1,2,3$, and $\nabla$ is the gradient associated with $g_{\mu\nu}$,
i.e., $\nabla =(m_1^{-1}\nabla_{\vq_1}, \dots,
m_N^{-1}\nabla_{\vq_N})$.}
\begin{equation}\label{Bohm}
    \frac{dQ}{dt} = v(Q)\,,\qquad
    v=\hbar \, \Im \, \frac{\Psi^* \nabla\Psi} {\Psi^* \, \Psi}\,.
\end{equation}
$\Psi=\Psi_t(q)$ is the wave function, which
evolves according to the Schr\"odinger equation
\begin{equation}\label{Seq}
    \I\hbar\frac{\partial\Psi}{\partial t} = H \Psi\,,
\end{equation}
with
\begin{equation}\label{Hamil}
     H=  -\frac{\hbar^2}{2} \Laplace + V
\end{equation}
for spinless particles, with $\Laplace = \div\nabla$. For particles
with spin, $\Psi$ takes values in the appropriate spin space $\CCC^k$,
$V$ may be matrix valued, and numerator and denominator of
\eqref{Bohm} have to be understood as involving inner products in spin
space. The secret of the success of Bohmian mechanics in yielding the
predictions of standard quantum mechanics is the fact that the
configuration $Q_t$ is $|\Psi_t|^2$-distributed in configuration space
at all times $t$, provided that the initial configuration $Q_0$ (part
of the Cauchy data of the theory) is so distributed.  This property,
called \emph{equivariance} in \cite{DGZ}, suffices for empirical
agreement between \emph{any} quantum theory (such as a QFT) and
\emph{any} version thereof with additional (often called ``hidden'')
variables $Q$, provided the outcomes of all experiments are registered
or recorded in these variables. That is why equivariance will be our
guide for obtaining the dynamics of the particles.

The equivariance of Bohmian mechanics follows immediately from comparing
the continuity equation for a probability distribution $\rho$
associated with (\ref{Bohm}),
\begin{equation}\label{master}
   \frac{\partial \rho}{\partial t} = -\div(\rho v)\,,
\end{equation}
with the equation satisfied by $|\Psi|^2$ which follows from
(\ref{Seq}),
\begin{equation}\label{continuity1}
   \frac{\partial |\Psi|^2}{\partial t}(q,t) = \frac{2}{\hbar} \, \Im
   \, \Big[ \Psi^*(q,t)\, (H\Psi)(q,t) \Big]\,.
\end{equation}
In fact, it follows from (\ref{Hamil}) that
\begin{equation}\label{JJJ}
   \frac{2}{\hbar} \, \Im \, \Big[ \Psi^*(q,t)\, (H\Psi)(q,t) \Big]=
   -\div\Big[\hbar \, \Im \, \Psi^*(q,t) \nabla\Psi(q,t)  \Big]
\end{equation}
so, recalling (\ref{Bohm}), one obtains that
\begin{equation}\label{continuity2}
   \frac{\partial |\Psi|^2}{\partial t} =  -\div(|\Psi|^2 v)\,,
\end{equation}
and hence that if $\rho_t=|\Psi_t|^2$ at some time $t$ then
$\rho_t=|\Psi_t|^2$ for \emph{all} times.  Equivariance is an
expression of the compatibility between the Schr\"odinger evolution
for the wave function and the law, such as (\ref{Bohm}), governing the
motion of the actual configuration.  In \cite{DGZ}, in which we were
concerned only with the Bohmian dynamics \eqref{Bohm}, we spoke of the
distribution $|\Psi|^2$ as being equivariant.  Here we wish to find
processes for which we have equivariance, and we shall therefore speak
of equivariant processes and motions.

\subsection{Equivariant Markov Processes}

The study of example QFTs like that of \cite{crea1} has lead us to the
consideration of Markov processes as candidates for the equivariant
motion of the configuration $Q$ for Hamiltonians $H$ more general than
those of the form \eqref{Hamil}.

Consider a Markov process $Q_t$ on configuration space.  The
transition probabilities are characterized by the \emph{backward
generator} $L_t$, a (time-dependent) linear operator acting on
functions $f$ on configuration space:
\begin{equation}\label{backgenerator}
   L_t f(q) = \frac{d}{ds} \EEE (f(Q_{t+s})|Q_t = q)
\end{equation}
where $d/ds$ means the right derivative at $s=0$ and
$\EEE(\,\cdot\,|\,\cdot\,)$ denotes the conditional expectation.
Equivalently, the transition probabilities are characterized by the
\emph{forward generator} $\generator_t$ (or, as we shall simply say,
\emph{generator}), which is also a linear operator but acts on
(signed) measures on the configuration space.  Its defining property
is that for every process $Q_t$ with the given transition
probabilities, the distribution $\rho_t$ of $Q_t$ evolves according to
\begin{equation}\label{rhoL}
   \frac{\partial \rho_t}{\partial t} = \generator_t \rho_t\,.
\end{equation}
$\generator_t$ is the dual of $L_t$ in the sense that
\begin{equation}\label{generatorduality}
   \int f(q) \, \generator_t \rho(dq) = \int L_t f(q) \, \rho(dq)\,.
\end{equation}
We will use both $L_t$ and $\generator_t$, whichever is more
convenient.  We will encounter several examples of generators in the
subsequent sections.

We can easily extend the notion of equivariance from deterministic to
Markov processes. Given the Markov transition probabilities, we say that
\emph{the $|\Psi|^2$ distribution is equivariant} if and only if for all
times $t$ and $t'$ with $t<t'$, a configuration $Q_t$ with distribution
$|\Psi_t|^2$ evolves, according to the transition probabilities, into a
configuration $Q_{t'}$ with distribution $|\Psi_{t'}|^2$. In this case, 
we
also simply say  that the transition probabilities are
\emph{equivariant}, without explicitly mentioning $|\Psi|^2$. 
Equivariance
is equivalent to
\begin{equation}\label{genequivariance}
   \generator_t |\Psi_t|^2 = \frac{\partial |\Psi_t|^2}{\partial t}
\end{equation}
for all $t$. When \eqref{genequivariance} holds (for a fixed $t$) we
also say that $\generator_t$ is an \emph{equivariant generator} (with
respect to $\Psi_t$ and $H$). Note that this definition of
equivariance agrees with the previous meaning for deterministic
processes.

We call a Markov process $Q$ \emph{equivariant} if and only if for every
$t$ the distribution $\rho_t$ of $Q_t$ equals $|\Psi_t|^2$. For this to 
be
the case, equivariant transition probabilities are necessary but not
sufficient. (While  for a Markov process $Q$  to have equivariant
transition probabilities amounts to the property that if $\rho_t =
|\Psi_t|^2$ for one time $t$, where $\rho_t$ denotes the distribution of
$Q_t$, then $\rho_{t'} = |\Psi_{t'}|^2$ for every $t'>t$,  according to
our definition of an equivariant Markov process, in fact $\rho_t =
|\Psi_t|^2$ for all $t$.)  However, for equivariant transition
probabilities there exists a unique equivariant Markov process.

The crucial idea for our construction of an equivariant Markov process
is to note that \eqref{continuity1} is completely general, and to find
a generator $\generator_t$ such that the right hand side of
(\ref{continuity1}) can be read as the action of $\generator$ on $\rho
= |\Psi|^2$,
\begin{equation}\label{mainequ}
    \frac{2}{\hbar} \, \Im \, \Psi^* H\Psi = \generator |\Psi|^2\,.
\end{equation}
We shall implement this idea beginning in Section \ref{sec:mini1},
after a review of jump processes and some general considerations. But
first we shall illustrate the idea with the familiar case of Bohmian
mechanics.

For $H$ of the form \eqref{Hamil}, we have (\ref{JJJ}) and hence that
\begin{equation}\label{mequ}
   \frac{2}{\hbar} \, \Im \, \Psi^*H\Psi = -\div\left(\hbar \, \Im \,
   \Psi^* \nabla\Psi \right) = -\div\left( |\Psi|^2 \hbar \, \Im \,
   \frac{\Psi^* \nabla\Psi} {|\Psi|^2} \right) \,.
\end{equation}
Since the generator of the (deterministic) Markov process
corresponding to the dynamical system $dQ/dt=v(Q)$ given by a velocity
vector field $v$ is
\begin{equation}\label{dynamical}
   \generator \rho = -\div(\rho v)\,,
\end{equation}
we may recognize the last term of (\ref{mequ}) as $\generator
|\Psi|^2$ with $\generator$ the generator of the deterministic process
defined by \eqref{Bohm}. Thus, as is well known, Bohmian mechanics
arises as the natural equivariant process on configuration space
associated with $H$ and $\Psi$.

To be sure, Bohmian mechanics is not the only solution of
(\ref{mainequ}) for $H$ given by \eqref{Hamil}. Among the alternatives
are Nelson's stochastic mechanics \cite{stochmech} and other velocity
formulas \cite{Deotto}. However, Bohmian mechanics is the most natural
choice, the one most likely to be relevant to physics. It is, in fact,
the canonical choice, in the sense of minimal process which we shall
explain in Section \ref{sec:mini}.

\subsection{Equivariant Jump Processes}\label{sec:revjump}

Let $\conf$ denote the configuration space of the process,
whatever sort of space that may be (vector space, lattice, manifold,
etc.); mathematically speaking, we need that $\conf$ be a measurable
space.  A (pure) jump process is a Markov process on $\conf$ for which
the only motion that occurs is via jumps. Given that $Q_t =q$, the
probability for a jump to $q'$, i.e., into the infinitesimal volume
$dq'$ about $q'$, by time $t+dt$ is $\sigma_t(dq'|q)\, dt$, where
$\sigma$ is called the \emph{jump rate}. In this notation, $\sigma$ is
a finite measure in the first variable; $\sigma(B|q)$ is the rate (the
probability per unit time) of jumping to somewhere in the set
$B\subseteq\conf$, given that the present location is $q$. The overall
jump rate is $\sigma(\conf|q)$.

It is often the case that $\conf$ is equipped with a distinguished
measure, which we shall denote by $dq$ or $dq'$, slightly abusing
notation.  For example, if $\conf = \RRR^d$, $dq$ may be the Lebesgue
measure, or if $\conf$ is a Riemannian manifold, $dq$ may be the
Riemannian volume element. When $\sigma(\,\cdot\,|q)$ is absolutely
continuous relative to the distinguished measure, we also write
$\sigma(q'|q)\, dq'$ instead of $\sigma(dq'|q)$.  Similarly, we
sometimes use the letter $\rho$ for denoting a measure and sometimes
the density of a measure, $\rho(dq) = \rho(q)\,dq$.

A jump first occurs when a random waiting time $T$ has elapsed, after 
the
time $t_0$ at which the process was started or at which the most
recent previous jump has occurred.  For purposes of simulating or
constructing the process, the destination $q'$ can be chosen at the
time of jumping, $t_0 + T$, with probability distribution
$\sigma_{t_0+T} (\conf|q)^{-1} \, \sigma_{t_0+T} (\,\cdot\,|q)$. In
case the overall jump rate is time-independent, $T$ is exponentially
distributed with mean $\sigma(\conf|q)^{-1}$. When the
rates are time-dependent---as they will typically be in what
follows---the waiting time remains such that
\[
   \int_{t_0}^{t_0+T} \sigma_t(\conf|q) \, dt
\]
is exponentially distributed with mean 1, i.e., $T$ becomes
exponential after a suitable (time-dependent) rescaling of time. For
more details about jump processes, see \cite{Breiman}.

The generator of a pure jump process can be expressed in terms of the
rates:
\begin{equation}\label{continuity3}
   \generator_\sigma \rho(dq) = \int\limits_{q'\in\conf}  \Big(
   \sigma(dq|q') \rho(dq') - \sigma(dq'|q) \rho(dq) \Big)\,,
\end{equation}
a ``balance'' or ``master'' equation expressing $\partial
\rho/\partial t$ as the gain due to jumps to $dq$ minus the loss due
to jumps away from $q$.

We shall say that jump rates $\sigma$ are \emph{equivariant} if
$\generator_\sigma$ is an equivariant generator.  It is one of our goals
in
this paper to describe  a general scheme for obtaining equivariant jump
rates. In Sections \ref{sec:mini1} and \ref{sec:mini2} we will explain 
how
this leads us to formula \eqref{tranrates}.

\subsection{Process Additivity}\label{sec:introadd}

The Hamiltonian of a QFT usually comes as a sum, such as
\begin{equation}\label{Hsum}
   H = H_0 + H_\inter
\end{equation}
with $H_0$ the free Hamiltonian and $H_\inter$ the interaction
Hamiltonian. If several particle species are involved, $H_0$ is itself
a sum containing one free Hamiltonian for each species. The left hand
side of (\ref{mainequ}), which should govern our choice of the
generator, is then also a sum,
\begin{equation}\label{Hsumgen}
   \frac{2}{\hbar} \, \Im \, \Psi^* H_0 \Psi + \frac{2}{\hbar} \, \Im
   \, \Psi^* H_\inter \Psi = \generator |\Psi|^2\,.
\end{equation}
This opens the possibility of finding a generator $\generator$ by
setting $\generator = \generator_0 + \generator_\inter$, provided we
have generators $\generator_0$ and $\generator_\inter$
corresponding to $H_0$ and $H_\inter$ in the sense that
\begin{subequations}
\begin{align}
   \frac{2}{\hbar} \, \Im \, \Psi^* H_0 \Psi
   &= \generator_0 |\Psi|^2 \\
   \frac{2}{\hbar} \, \Im \, \Psi^* H_\inter \Psi
   &= \generator_\inter |\Psi|^2\,.
\end{align}
\end{subequations}
This feature of (\ref{mainequ}) we call \emph{process additivity}; it
is based on the fact that the left hand side of (\ref{mainequ}) is
linear in $H$.  Note that the backward generator of the process with
forward generator $\generator_0 + \generator_\inter$ is $L_0 +
L_\inter$; thus forward and backward generators lead to the same
notion of process additivity, and to the same process corresponding to
$H_0 + H_\inter$.  In many cases, as will be elaborated in Section
\ref{sec:free}, $H_0$ is based on an operator known from quantum
mechanics (e.g., the Dirac operator), in such a way that
$\generator_0$ can be obtained from the appropriate Bohmian law of
motion. In Section \ref{sec:mini1} we will explain how
$\generator_\inter$ can usually be taken as the generator of a jump
process.

Our proposal is to take seriously the process generated by $\generator
= \generator_0 + \generator_\inter$ and regard it as the process
naturally associated with $H$. The bottom line is that process
additivity provides a \emph{method of constructing} a Bell-type
theory.

Obviously, the mathematical observation of process additivity (that
sums of generators define an equivariant process associated with sums
of Hamiltonians) applies not only to the splitting of $H$ into a free
and an interaction contribution, but to every case where $H$ is a sum.
And it seems  that process additivity provides a physically very
reasonable process in every case where $H$ is naturally a sum, in fact
the most reasonable process: the one that should be considered
\emph{the} Bell-type process, defining \emph{the} Bell-type theory.

\subsection{What Added Processes May Look Like}

To get some feeling for what addition of generators, $\generator =
\generator_1 + \generator_2$, means for the corresponding processes,
we consider some examples. First consider two deterministic processes
(on the same configuration space), having generators of the form
$\generator \rho = -\div(\rho v)$.  To add the generators obviously
means to add the velocity vector fields, $v=v_1 + v_2$, so the
resulting velocity is a superposition of two contributions.

Next consider a pure jump process. Since, according to
(\ref{continuity3}), the generator $\generator$ is linear in $\sigma$,
adding generators means adding rates, $\sigma = \sigma_1 +
\sigma_2$. This is equivalent to saying there are two kinds of jumps:
if the present location is $q\in\conf$, with probability
$\sigma_1(\conf|q)\,dt$ the process performs a jump of the first type
within the next $dt$ time units, and with probability
$\sigma_2(\conf|q)\,dt$ a jump of the second type. That does not mean,
however, that one can decide from a given realization of the process
which jump was of which type.

Next suppose we add the generators of a deterministic and a jump 
process,
\begin{equation}\label{continuity4}
   \generator \rho(q) = -\div(\rho v)(q) + \int\limits_{q'\in\conf}
   \Big( \sigma(q|q')\, \rho(q') - \sigma(q'|q)\, \rho(q) \Big) dq'\,.
\end{equation}
This process moves with velocity $v(q)$ until it jumps to $q'$, where
it continues moving, with velocity $v(q')$. The jump rate may vary
with time in two ways: first because $\sigma$ may be time-dependent,
second because $\sigma$ may be position-dependent and $Q_t$ moves with
velocity $v$. One can easily understand (\ref{continuity4}) in terms
of gain or loss of probability density due to motion and jumps. So
this process is piecewise deterministic: although the temporal length
of the pieces (the intervals between two subsequent jumps) and the
starting points (the jump destinations) are random, given this data
the trajectory is determined.

The generator of the Wiener process in $\RRR^d$ is the Laplacian, and
to add to it the generator of a deterministic process means to
introduce a drift. Note that this is different from adding, in
$\RRR^d$, a Wiener process to a solution of the deterministic
process. In spaces like $\RRR^d$, where it so happens that one is
allowed to add locations, there is a danger of confusing addition of
generators with addition of realizations. Whenever we speak of adding
processes, it means we add generators.

To add generators of a diffusion and a pure jump process yields what
is often called a jump diffusion process, one making jumps with time-
and position-dependent rates and following a diffusion path in between.
Diffusion processes, however, will play almost no role in this paper.

\subsection{Integral Operators Correspond to Jump Processes}
\label{sec:mini1}

We now address the interaction part $H_\inter$ of the Hamiltonian
(\ref{Hsum}).  In QFTs with cutoffs it is usually the case that
$H_\inter$ is an integral operator.  For that reason, we shall in this
work focus on integral operators for $H_\inter$.  We now point out why
the naturally associated process is a pure jump process.  For short,
we will write $H$ rather than $H_\inter$ in this and the subsequent
section. For the time being, think of $\conf$ as $\RRR^d$ and of wave
functions as complex valued.

What characterizes jump processes versus continuous processes is that
some amount of probability that vanishes at $q\in\conf$ can reappear
in an entirely different region of configuration space, say at
$q'\in\conf$. This is manifest in the equation for $\partial
\rho/\partial t$, (\ref{continuity3}): the first term in the integrand
is the probability increase due to arriving jumps, the second the
decrease due to departing jumps, and the integration over $q'$
reflects that $q'$ can be anywhere in $\conf$. This suggests that
Hamiltonians for which the expression \eqref{continuity1} for
$\partial |\Psi|^2/\partial t$ is naturally an integral over $dq'$
correspond to pure jump processes. So when is the left hand side of
(\ref{mainequ}) an integral over $dq'$?  When $H$ is an integral
operator, i.e., when $\sp{q}{H|q'}$ is not merely a formal symbol, but
represents an integral kernel that exists as a function or a measure and
satisfies
\begin{equation}
   (H\Psi)(q) = \int dq'\,\sp{q}{H|q'}\, \Psi(q')\,.
\end{equation}
In this case, we should choose the jump rates in such a way that,
when $\rho = |\Psi|^2$,
\begin{equation}\label{la1}
   \sigma(q|q') \,\rho(q') - \sigma(q'|q) \,\rho(q) = \frac{2}{\hbar}
   \, \Im \, \Psi^*(q)\, \sp{q}{H|q'} \, \Psi(q') \,,
\end{equation}
and this suggests, since jump rates must be nonnegative (and the right
hand side of \eqref{la1} is anti-symmetric), that
\[
   \sigma(q|q') \,\rho(q') = \Big[ \frac{2}{\hbar} \, \Im \,
   \Psi^*(q)\, \sp{q}{H|q'} \, \Psi(q') \Big]^+
\]
(where $x^+$ denotes the positive part of $x\in\RRR$, that is, $x^+$ is
equal to $x $ for $x>0$ and is zero otherwise), or
\begin{equation}\label{mini1}
   \sigma(q|q') = \frac{ \big[ (2/\hbar) \, \Im \, \Psi^*(q) \, \sp{q}
   {H|q'} \, \Psi(q') \big]^+}{\Psi^*(q')\, \Psi(q')} .
\end{equation}
These rates are an instance of what we call the \emph{minimal jump
rates} associated with $H$ (and $\Psi$).  The name comes from the fact
that they are actually the minimal possible values given (\ref{la1}),
as is expressed by the inequality \eqref{minimality} and will be
explained in detail in Section \ref{sec:mini4}.  Minimality entails
that at any time $t$, one of the transitions $q_1 \to q_2$ or $q_2 \to
q_1$ is forbidden. We will call the process defined by the minimal
jump rates the \emph{minimal jump process} (associated with $H$).

In contrast to jump processes, continuous motion, as in Bohmian
mechanics, corresponds to such Hamiltonians that the formal matrix
elements $\sp{q}{H|q'}$ are nonzero only infinitesimally close to the
diagonal, and in particular to differential operators like the
Schr\"odinger Hamiltonian (\ref{Hamil}), which has matrix elements of
the type $\delta''(q-q') + V(q) \,\delta(q-q')$. We can summarize the
situation, as a rule of thumb, by the following table:

\begin{center}
\begin{tabular}{|r|l|}
   \hline
   A contribution to $H$ that is a \ldots & corresponds to \ldots\\\hline
   integral operator & jumps\\
   differential operator & deterministic continuous motion\\
   multiplication operator & no motion ($\generator = 0$)\\\hline
\end{tabular}
\end{center}

The minimal jump rates as given by (\ref{mini1}) have some nice
features.  The possible jumps for this process correspond to the
nonvanishing matrix elements of $H$ (though, depending on the state
$\Psi$, even some of the jump rates corresponding to nonvanishing
matrix elements of $H$ might happen to vanish).  Moreover, in their
dependence on the state $\Psi$, the jump rates $\sigma$ depend only
``locally'' upon $\Psi$: the jump rate for a given jump $q'\to q$
depends only on the values $\Psi(q')$ and $\Psi(q)$ corresponding to
the configurations linked by that jump.  Discretizing $\RRR^3$ to a
lattice $\varepsilon \ZZZ^3$, one can obtain Bohmian mechanics as a
limit $\varepsilon\to 0$ of minimal jump processes
\cite{Sudbery,Vink}, whereas greater-than-minimal jump rates lead to
Nelson's stochastic mechanics \cite{stochmech} and similar diffusions,
such as (\ref{diffusion}); see \cite{Vink,Guerra}. If the
Schr\"odinger operator \eqref{Hamil} is approximated in other ways by
operators corresponding to jump processes, e.g., by $H_\varepsilon =
\E^{-\varepsilon H} H \E^{-\varepsilon H}$, the minimal jump processes
presumably also converge to Bohmian mechanics.

We have reason to believe that there are lots of self-adjoint
operators which do not correspond to any stochastic process that can
be regarded as defined, in any reasonable sense, by
\eqref{mini1}.\footnote{Consider, for example, $H = p \cos p$ where
$p$ is the one-dimensional momentum operator $-\I\hbar
\partial/\partial q$. Its formal kernel $\sp{q}{H|q'}$ is the
distribution $-\frac{\I}{2} \delta'(q-q'-1) - \frac{\I}{2}
\delta'(q-q'+1)$, for which \eqref{mini1} would not have a meaning.
{}From a sequence of smooth functions converging to this distribution,
one can obtain a sequence of jump processes with rates \eqref{mini1}:
the jumps occur very frequently, and are by amounts of approximately
$\pm 1$.  A limiting process, however, does not exist.}  But such
operators seem never to occur in QFT. (The Klein--Gordon operator
$\sqrt{m^2 c^4 - \hbar^2 c^2 \Laplace}$ does seem to have a process,
but it requires a more detailed discussion which will be provided in a
forthcoming work \cite{klein2}.)

\subsection{Minimal Jump Rates}
\label{sec:mini2}

The reasoning of the previous section applies to a far more general
setting than just considered: to arbitrary configuration spaces
$\conf$ and ``generalized observables''---POVMs---defining, for our
purposes, what the ``position representation'' is. We now present this
more general reasoning, which leads to one of the main formulas of
this paper, (\ref{tranrates}).

The process we construct relies on the following ingredients from QFT:
\begin{enumerate}
\item A Hilbert space $\Hilbert$ with scalar product $\sp{\Psi}
   {\Phi}$.

\item A unitary one-parameter group $U_t$ in $\Hilbert$ with
   Hamiltonian $H$,
   \[
     U_t = \E^{-\frac{\I}{\hbar}tH}\,,
   \]
   so that in the Schr\"odinger picture the state $\Psi$ evolves
   according to
   \begin{equation}
     \I\hbar\frac{d\Psi_t}{dt} = H\Psi_t\,.
   \end{equation}
   $U_t$ could be part of a representation of the Poincar\'e group.

\item A positive-operator-valued measure (POVM) $\pov(dq)$ on $\conf$
   acting on $\Hilbert$, so that the probability that the system in the
   state $\Psi$ is localized in $dq$ at time $t$ is
   \begin{equation} \label{mis}
     \measure_t(dq)= \sp{\Psi_t}{\pov(dq)| \Psi_t} \,.
   \end{equation}
\end{enumerate}

Mathematically, a POVM $\pov$ on $\conf$ is a countably additive set
function (``measure''), defined on measurable subsets of $\conf$, with
values in the positive (bounded self-adjoint) operators on (a Hilbert
space) $\Hilbert$, such that $\pov(\conf)$ is the identity
operator.\footnote{The countable additivity is to be understood as in
the sense of the weak operator topology. This in fact implies that
countable additivity also holds in the strong topology.}  Physically,
for our purposes, $\pov(\,\cdot\,)$ represents the (generalized)
position observable, with values in $\conf$.  The notion of POVM
generalizes the more familiar situation of observables given by a set
of commuting self-adjoint operators, corresponding, by means of the
spectral theorem, to a projection-valued measure (PVM): the case where
the positive operators are projection operators.  A typical example is
the single Dirac particle: the position operators on
$L^2(\RRR^3,\CCC^4)$ induce there a natural PVM $\pov_0(\,\cdot\,)$:
for any Borel set $B\subseteq \RRR^3$, $\pov_0(B)$ is the projection
to the subspace of functions that vanish outside $B$, or,
equivalently, $\pov_0(B)\Psi(q) = \1_B(q) \, \Psi(q)$ with $\1_B$ the
indicator function of the set $B$.  Thus, $\sp{\Psi} {\pov_0 (dq)|
\Psi} = |\Psi(q)|^2 dq$.  When one considers as Hilbert space
$\Hilbert$ only the subspace of positive energy states, however, the
localization probability is given by $\pov(\,\cdot\,) = P_+
\pov_0(\,\cdot\,) I$ with $P_+:L^2(\RRR^3,\CCC^4) \to \Hilbert$ the
projection and $I:\Hilbert \to L^2(\RRR^3,\CCC^4)$ the inclusion
mapping. Since $P_+$ does not commute with most of the operators
$\pov_0(B)$, $\pov (\,\cdot\,)$ is no longer a PVM but a genuine
POVM\footnote{This situation is indeed more general than it may seem.
By a theorem of Naimark \cite[p.~142]{Davies}, every POVM $\pov
(\,\cdot\,)$ acting on $\Hilbert$ is of the form $\pov(\,\cdot\,) =
P_+ \pov_0 (\,\cdot\,) P_+$ where $\pov_0$ is a PVM on a larger
Hilbert space, and $P_+$ the projection to
$\Hilbert$. \label{ft:Naimark}} and consequently does not correspond
to any position operator---although it remains true (for $\Psi$ in the
positive energy subspace) that $\sp{\Psi}{\pov(dq)| \Psi} =
|\Psi(q)|^2 dq$.  That is why in QFT, the position observable is
indeed more often a POVM than a PVM. POVMs are also relevant to
photons \cite{ali,kraus}.  In one approach, the photon wave function
$\Psi: \RRR^3 \to \CCC^3$ is subject to the constraint condition
$\nabla \cdot \Psi = \partial_1 \Psi_1 + \partial_2 \Psi_2 +
\partial_3 \Psi_3 =0$. Thus, the physical Hilbert space $\Hilbert$ is
the (closure of the) subspace of $L^2(\RRR^3,\CCC^3)$ defined by this
constraint, and the natural PVM on $L^2(\RRR^3,\CCC^3)$ gives rise, by
projection, to a POVM on $\Hilbert$.  So much for POVMs. Let us get
back to the construction of a jump process.

The goal is to specify equivariant jump rates $\sigma = \sigma^{\Psi, H,
\pov}$, i.e., such rates that
\begin{equation}\label{equirates}
   \generator_\sigma \measure = \frac{d\measure}{dt} \,.
\end{equation}
To this end, one may take the following steps:

\begin{enumerate}
\item Note that
   \begin{equation}\label{dPdt}
     \frac{d\measure_t(dq)}{dt} = \frac{2}{\hbar} \, \Im \,
     \sp{\Psi_t}{\pov(dq) H| \Psi_t}\,.
   \end{equation}
\item Insert the resolution of the identity $I = 
\int\limits_{q'\in\conf}
   \pov(dq')$ and obtain
   \begin{equation}\label{dPdtJ}
     \frac{d\measure_t(dq)}{dt} =\int\limits_{q'\in\conf}
     \current_t(dq,dq') \,,
   \end{equation}
   where
   \begin{equation}\label{Jdef}
     \current_t(dq,dq') = \frac{2}{\hbar} \,
     \Im \, \sp{\Psi_t}{\pov(dq)H \pov(dq')| \Psi_t} \,.
   \end{equation}
\item Observe that $\current$ is anti-symmetric, $\current(dq',dq) = -
   \current(dq,dq')$.  Thus, since $x = x^+ - (-x)^+$,
   \[
     \current(dq,dq') = \left[(2/\hbar) \, \Im \, \sp{\Psi} {\pov(dq) H
     \pov(dq') |\Psi}\right]^+ - \left[(2/\hbar)\, \Im \, \sp{\Psi}
     {\pov(dq') H \pov(dq) |\Psi}\right]^+ .
   \]
\item Multiply and divide both terms by $\measure(\,\cdot\,)$,
   obtaining that
   \begin{eqnarray*}
     \int\limits_{q'\in\conf} \current(dq,dq') = \int\limits_{q'\in\conf}
     \bigg( \hspace{-3ex} &&
     \frac{[(2/\hbar) \, \Im \, \sp{\Psi} {\pov(dq) H \pov(dq')| 
\Psi}]^+}
     {\sp{\Psi}{\pov(dq')| \Psi}} \measure(dq') -
   \\-&&
     \frac{[(2/\hbar) \, \Im \, \sp{\Psi} {\pov(dq') H \pov(dq)| \Psi}
     ]^+} {\sp{\Psi} {\pov(dq)| \Psi}} \measure(dq) \bigg) \,.
\end{eqnarray*}
\item By comparison with \eqref{continuity3}, recognize the right hand
   side of the above equation as $\generator_\sigma \measure$, with
   $\generator_\sigma$ the generator of a Markov jump process with jump
   rates
   \begin{equation} \label{tranrates}
     \sigma(dq|q')= \frac{[(2/\hbar) \, \Im \, \sp{\Psi} {\pov(dq) H
     \pov(dq')| \Psi}]^+}{\sp{\Psi}{\pov(dq')| \Psi}}\,,
\end{equation}
    which we call the \emph{minimal jump rates}.
\end{enumerate}
Mathematically, the right hand side of this formula as a function of
$q'$ must be understood as a density (Radon--Nikod{\'y}m derivative)
of one measure relative to another.\footnote{Quite aside from the
previous discussion, it is perhaps worth noting that there are not so
many expressions in $H,\pov$, and $\Psi$ that would meet the formal
criteria for being a candidate for the jump rate. Since the only
connection between abstract Hilbert space and configuration space is
by $\pov$, which leads to \emph{measures} on $\conf$, the only way to
obtain a \emph{function} on $\conf$ is to form a Radon--Nikod{\'y}m
quotient of two measures, $\sigma(q') = A(dq')/B(dq')$. Since $\sigma$
must be a measure-valued function, the numerator should be a
bi-measure (a measure in each of two variables). The simplest measure
one can form from $H,\pov$, and $\Psi$ is $\sp{\Psi}{\pov(dq)|\Psi}$;
the simplest bi-measures are $\sp{\Psi}{H^{n_1} \pov(dq) H^{n_2}
\pov(dq') H^{n_3}| \Psi}$. Jump rates must have dimension 1/time, and
the only object at hand having this dimension is $H/\hbar$. Thus, $H$
can appear only once in the numerator. The expressions $\sp{\Psi}{H
\pov(dq) \pov(dq')| \Psi}$ and $\sp{\Psi}{\pov(dq) \pov(dq') H| \Psi}$
are no good because for PVMs $\pov$ they are concentrated on the
diagonal of $\conf \times \conf$ and hence do not lead to nontrivial
jumps. Let us write $\mu$ for the measure-valued function we have
arrived at:
\[
   \mu (dq,q') = \frac{1}{\hbar} \frac{\sp{\Psi}{\pov(dq) H \pov(dq')
   | \Psi}} {\sp{\Psi}{\pov(dq')|\Psi}}\,.
\]
This provides \emph{complex} measures, whereas $\sigma(\,\cdot\,|q')$
must be a positive real measure. There are not many ways of forming
a positive real measure from a complex one, the essential ones being
\[
   |\mu|, |\Re \, \mu|, |\Im \, \mu|, (\Re \, \mu)^+, (\Re \, \mu)^-,
   (\Im \, \mu)^+, (\Im \, \mu)^-
\]
times a numerical constant $\lambda>0$. One could of course form
additional expressions at the price of higher complexity.

This has gotten us already pretty close to the minimal rates
\eqref{tranrates}, which correspond to $\sigma = 2(\Im \, \mu)^+$. To
proceed further, we might demand the absence of unnecessary jumps;
that means that at any time, either the jump $q_1 \to q_2$ or $q_2 \to
q_1$ is forbidden; this leaves only $\lambda (\Im \,
\mu)^\pm$. Moreover, $2 (\Im \, \mu)^+$ is the only expression in the
list that has Bohmian mechanics as a limiting case or implies
equivariance. Furthermore it corresponds to the natural guess
\eqref{Ltilde} for a backward generator, discussed in Section
\ref{sec:mini}.} The plus symbol denotes the positive part of a signed
measure; it can also be understood as applying the plus function, $x^+
= \max (x,0)$, to the density, if it exists, of the numerator.

To sum up, we have argued that with $H$ and $\Psi$ is naturally
associated a Markov jump process $Q_t$ whose marginal distributions
coincide at all times by construction with the quantum probability
measure, $\rho_t(\,\cdot\,) = \measure_t(\,\cdot\,)$, so that $Q_t$ is
an equivariant Markov process.

In Section~4 of \cite{crea2A}, we establish precise conditions on
$H,\pov$, and $\Psi$ under which the jump rates \eqref{tranrates} are
well-defined and finite $\measure$-almost everywhere, and prove that
in this case the rates are equivariant, as suggested by the steps 1-5
above. It is perhaps worth remarking at this point that any $H$ can be
approximated by Hamiltonians $H_n$ (namely Hilbert--Schmidt operators)
for which the rates \eqref{tranrates} are always (for all $\Psi$)
well-defined and equivariant \cite{crea2A}. Concerning this, see also
the end of Section \ref{sec:mini}.

\subsection{Process Associated with the Free Hamiltonian}
\label{sec:free}

We now address the free Hamiltonian $H_0$ of a QFT.  We describe the
process naturally associated with $H_0$, when this is the second
quantized Schr\"odinger or Dirac operator.  We will treat more general
free Hamiltonians in the next section. We shall consider here only
Hamiltonians for one type of particle.

We first define the configuration space $\conf$.  Let us write
$\conf^{(1)}$ (``one-particle configuration space'') for physical
space; this is typically, but not necessarily, $\RRR^3$.  The space
$\conf$ in which the ``free process'' takes place is the configuration
space for a variable number of identical particles; we call it $\Gamma
\conf^{(1)}$.  It can be defined as the space of all finite
subsets-with-multiplicities of $\conf^{(1)}$.  A
set-with-multiplicities consists of a set and, for each element $x$ of
the set, a positive integer, called the multiplicity of $x$. The
number of particles in a configuration $q$ is the sum of its
multiplicities, $\#q$. Such configurations describe several identical
particles, some of which may be located at the same position in
space. Equivalently, one could say that $\Gamma \conf^{(1)}$ is the
set of all mappings $n:\conf^{(1)} \to \NNN\cup\{0\}$ (meaning the
number of particles at a given location) such that
\[
   \sum_{\vq \in \conf^{(1)}} n(\vq ) < \infty\,.
\]
Another equivalent definition is the set of all finite nonnegative
measures $n(\,\cdot\,)$ on $\conf^{(1)}$ that assume only integer
values; the meaning of $n(R)$ is the number of particles in the region
$R$ of physical space. Finally, one can define
\[
   \Gamma \conf^{(1)} = \bigcup_{n=0}^\infty \conf^{(n)} \mbox{ where }
   \conf^{(n)} = (\conf^{(1)})^n/\mbox{permutations}.
\]

A related space, for which we write $\Gommo \conf^{(1)}$, is the space
of all finite subsets of $\conf^{(1)}$; it is contained in $\Gamma
\conf^{(1)}$, after obvious identifications. In fact, $\Gommo
\conf^{(1)} = \Gamma \conf^{(1)} \setminus \Delta$, where $\Delta$ is
the set of coincidence configurations, i.e., those having two or more
particles at the same position.  $\Gommo \conf^{(1)}$ is the union of
the spaces ${\conf}^{(n)}_{\neq}$ for $n=0,1,2, \ldots$, where
${\conf}^{(n)}_{\neq}$ is the space of subsets of $\conf^{(1)}$ with
$n$ elements.

For $\conf^{(1)} = \RRR^d$, the $n$-particle sector
${\conf}^{(n)}_{\neq}$ is a manifold of dimension $nd$ (see
\cite{identical} for a discussion of Bohmian mechanics on this
manifold).  If $d\geq 2$, the set $\Delta$ of coincidence
configurations has codimension $\geq 2$ and thus can usually be
ignored.  We can then replace $\Gamma \RRR^d$ by the somewhat simpler
space $\Gommo \RRR^d$.

The position POVM $\pov^{(1)}$ on $\conf^{(1)}$ (acting on the
one-particle Hilbert space) naturally leads to a POVM we call $\Gamma
\pov^{(1)}$ on $\conf = \Gamma \conf^{(1)}$, acting on Fock space (see
Section \ref{sec:GammaPOVM} for the definition).\footnote{The
coincidence configurations form a null set, $\Gamma \pov^{(1)}(\Delta)
=0$, when $\conf^{(1)}$ is a continuum, or, more precisely, when
$\pov^{(1)}$ is nonatomic as a measure.}  Since a configuration from
$\Gamma(\RRR^3)$ defines the number of particles and their positions,
the name ``position observable'' for $\pov = \Gamma \pov^{(1)}$
stretches the meaning of ``position'' somewhat: it now also
encompasses the number of particles.

We now give a description of the free process associated with the
second-quantized Schr\"odinger operator; it arises from Bohmian
mechanics.  Fock space $\Hilbert = \Fock$ is a direct sum
\begin{equation}\label{fockspace}
   \Fock= \bigoplus_{n=0}^{\infty} \Fock^{(n)} ,
\end{equation}
where $\Fock^{(n)}$ is the $n$-particle Hilbert space.  $\Fock^{(n)}$
is the subspace of symmetric (for bosons) or anti-symmetric (for
fermions) functions in $L^2 (\RRR^{3n}, (\CCC^{2s+1})^{\otimes n})$
for spin-$s$ particles. Thus, $\Psi \in \Fock$ can be decomposed into
a sequence $\Psi = \left( \Psi^{(0)}, \Psi^{(1)}, \ldots, \Psi^{(n)},
\ldots \right)$, the $n$-th member $\Psi^{(n)}$ being an $n$-particle
wave function, the wave function representing the $n$-particle sector
of the quantum state vector.  The obvious way to obtain a process on
$\conf = \Gamma \RRR^3$ is to let the configuration $Q(t)$, containing
$N = \#Q(t)$ particles, move according to the $N$-particle version of
Bohm's law (\ref{Bohm}), guided by $\Psi^{(N)}$.\footnote{As defined,
configurations are unordered, whereas we have written Bohm's law
\eqref{Bohm} for ordered configurations.  Thanks to the
(anti\nobreakdash-)symmetry of the wave function, however, all
orderings will lead to the same particle motion. For more about such
considerations, see our forthcoming work \cite{identical}.}  This is
indeed an equivariant process since $H_0$ has a block diagonal form
with respect to the decomposition (\ref{fockspace}),
\[
   H_0 = \bigoplus_{n=0}^\infty H_0^{(n)}\,,
\]
and $H_0^{(n)}$ is just a Schr\"odinger operator for $n$
noninteracting particles, for which, as we already know, Bohmian
mechanics is equivariant.  We used a very similar process in
\cite{crea1} (the only difference being that particles were numbered
in \cite{crea1}).

Similarly, if $H_0$ is the second quantized Dirac operator, we let a
configuration $Q$ with $N$ particles move according to the usual
$N$-particle Bohm--Dirac law \cite[p.~274]{BH}
\begin{equation}\label{BohmDirac}
   \frac{dQ}{dt} = c\frac{\Psi^*(Q) \, \alpha_{N} \, \Psi(Q)}
   {\Psi^*(Q) \, \Psi(Q)}
\end{equation}
where $c$ denotes the speed of light and $\alpha_{N} = (\valpha^{(1)},
\ldots, \valpha^{(N)})$ with $\valpha^{(k)}$ acting on the spin index
of the $k$-th particle.

\subsection{Other Approaches to the Free Process}
\label{sec:free2}

We will give below a general velocity formula, applicable to a wider
class of free Hamiltonians.  Alternatively, we can provide a free
process for any $H_0$ if we are given an equivariant process for the
one-particle Hamiltonian $H^{(1)}$.  This is based on the particular
mathematical structure of $H_0$, which can be expressed by saying it
arises from a one-particle Hamiltonian $H^{(1)}$ by applying a
``second quantization functor $\Gamma$'' \cite{RS}. That is, there is
an algorithm (in a bosonic or fermionic version) for forming, from a
one-particle Hilbert space $\Hilbert^{(1)}$ and a one-particle
Hamiltonian $H^{(1)}$, a Fock space $\Fock = \Gamma\Hilbert^{(1)}$ and
free Hamiltonian $H_0 = \Gamma H^{(1)}$.  And parallel to this
``second quantization'' algorithm, there is an algorithm for the
canonical construction, from a given equivariant one-particle Markov
process $Q^{(1)}_t$, of a process we call $\Gamma Q^{(1)}_t$ that
takes place in $\conf = \Gamma \conf^{(1)}$ and is equivariant with
respect to $H_0$.  This algorithm may be called the ``second
quantization'' of a Markov process.

The algorithm is described in Section \ref{sec:Gamma}.  What the
algorithm does is essentially to construct an $n$-particle version of
$Q^{(1)}_t$ for every $n$, and finally combine these by means of a
random particle number $N = N(t) = \# Q(t)$ which is constant under
the free process, parallel to the fact that the particle number
operator is conserved by $H_0$. We note further that the process
$\Gamma Q^{(1)}_t$ is deterministic if $Q^{(1)}_t$ is.  If we take the
one-particle process to be Bohmian mechanics or the Bohm--Dirac
motion, the algorithm reproduces the processes described in the
previous section.

The algorithm leaves us with the task of finding a suitable
one-particle law, which we do not address in this paper. For some
Hamiltonians, such as the Dirac operator, this is immediate, for
others it is rather nontrivial, or even unsolved. The Klein--Gordon
operator $\sqrt{m^2c^4 - \hbar^2c^2\Laplace}$ will be discussed in
forthcoming work \cite{klein2}, and for a study of photons see
\cite{photon}.

When $H_0$ is made of differential operators of up to second order
(which includes of course the Schr\"odinger and Dirac operators),
there is another way to characterize the process associated with
$H_0$, a way which allows a particularly succinct description of the
process and a particularly direct derivation and construction.  In
fact, we give a formula for its backward generator $L_0$, or
alternatively the velocity (or the forward generator $\generator_0$),
in terms of $H_0,\pov$, and $\Psi$.

We begin by defining, for any $H,\pov$, and $\Psi$, an operator
$L$ acting on functions $f:\conf \to \RRR$, which may or may
not be the backward generator of a process, by
\begin{equation}\label{LH}
   Lf(q) = \Re \frac{\sp{\Psi} {\pov(dq) \hat{L} \hat{f} |\Psi}}
   {\sp{\Psi} {\pov(dq)|\Psi}} = \Re \frac{\sp{\Psi} {\pov(dq)
   \frac{\I}{\hbar} [H,\hat{f}] |\Psi}} {\sp{\Psi} {\pov(dq) |\Psi}}.
\end{equation}
where $[\;,\,]$ means the commutator,
\begin{equation}\label{hatf}
   \hat{f} = \int\limits_{q \in \conf} f(q) \, \pov(dq)\,,
\end{equation}
and $\hat{L}$ is the ``generator'' of the (Heisenberg) time evolution of
the operator $\hat{f}$,
\begin{equation}\label{hatLdef}
   \hat{L}\hat{f} = \frac{d}{d\tau} \E^{\I H \tau/\hbar} \, \hat{f} \,
   \E^{-\I H \tau/\hbar} \Big|_{\tau =0} = \tfrac{\I}{\hbar}
   [H,\hat{f}] \,.
\end{equation}
(If $\pov$ is a PVM, then $\hat{f} = f(\hat{q})$, where $\hat{q}$ is
the configuration operator.)  \eqref{LH} could be guessed in the
following way: since $Lf$ is in a certain sense, see
\eqref{backgenerator}, the time derivative of $f$, it might be
expected to be related to $\hat{L} \hat{f}$, which is in a certain
sense, see \eqref{hatLdef}, the time derivative of $\hat{f}$.  As a
way of turning the operator $\hat{L} \hat{f}$ into a function $Lf(q)$,
the middle term in \eqref{LH} is an obvious possibility.  Note that
this way of arriving at \eqref{LH} does not make use of equivariance;
for another way that does, see Section \ref{sec:freeflow}.

The formula for the forward generator equivalent to \eqref{LH} reads
\begin{equation}\label{genH}
   \generator \rho(dq) = \Re \, \sp{\Psi}{\widehat{\tfrac{d\rho}
   {d\measure}}\, \tfrac{\I}{\hbar} [H, \pov (dq)] |\Psi},
\end{equation}
as follows from \eqref{generatorduality}.

Whenever $L$ is indeed a backward generator, we call it the
\emph{minimal free (backward) generator} associated with $\Psi, H$,
and $\pov$. (The name is based on the concept of minimal process as
explained in Section \ref{sec:mini}.) Then the corresponding process
is equivariant (see Section \ref{sec:freeflow}).  This is the case if
(and, there is reason to expect, \emph{only if}) $\pov$ is a PVM and
$H$ is a differential operator of up to second order in the position
representation, in which $\pov$ is diagonal.  In that case, the
process is deterministic, and the backward generator has the form $L =
v \cdot \nabla$ where $v$ is the velocity vector field; thus,
\eqref{LH} directly specifies the velocity, in the form of a
first-order differential operator $v \cdot \nabla$.  In case $H$ is
the $N$-particle Schr\"odinger operator with or without spin,
\eqref{LH} yields the Bohmian velocity \eqref{Bohm}, and if $H$ is the
Dirac operator, the Bohm--Dirac velocity \eqref{BohmDirac}.  To sum
up, in some cases definition \eqref{LH} leads to just the right
backward generator.

To return to our starting point: if the one-particle generator
$\generator^{(1)}$ arises from the one-particle Hamiltonian $H^{(1)}$
by \eqref{genH}, then \eqref{genH} also holds between the free generator
$\generator_0 = \Gamma \generator^{(1)}$ and the free Hamiltonian $H_0
= \Gamma H^{(1)}$.  (See Section \ref{sec:freeflow} for details.) In
other words, \eqref{LH} is compatible with the ``second quantization''
algorithm.  Thus, in relevant cases \eqref{LH} allows a direct
definition of the free process in terms of $H_0$, just as
\eqref{tranrates} directly defines, in terms of $H_\inter$, the jump
rates.

A relevant point is that the ``second quantization'' of a differential
operator is again a differential operator, in a suitable sense, and
has the same order.  Note also that \eqref{LH}, when applied to the
second quantized Schr\"odinger or Dirac Hamiltonian, defines the same
vector field on $\Gamma(\RRR^3)$ as described in the previous section.

\subsection{Bell-Type QFT}

We briefly summarize what we have obtained. A Bell-type QFT is about
particles moving in physical 3-space; their number and positions are
represented by a point $Q_t$ in configuration space $\conf$.  Provided
physical space is $\RRR^3$, $\conf$ is usually $\Gamma \RRR^3$ or a
Cartesian product of several such spaces, each factor representing a
different particle species.  $Q_t$ follows a Markov process in
$\conf$, which is governed by a state vector $\Psi$ in a suitable
Hilbert space $\Hilbert$.  $\Hilbert$ is related to $\conf$ by means
of a PVM or POVM $\pov$.  $\Psi$ undergoes a unitary evolution with
Hamiltonian $H$.  The process $Q_t$ usually consists of deterministic
continuous trajectories interrupted by stochastic jumps; more
generally, it arises by process additivity (i.e., by adding
generators) from a free process associated with $H_0$ and a jump
process associated with $H_\inter$.  The jump rates are given by
\eqref{tranrates} for $H= H_\inter$.  The free process arises from
Bohmian mechanics, or a suitable analogue, by a construction that can
be formalized as the ``second quantization'' of a one-particle Markov
process; when appropriate, it is defined directly by \eqref{LH}.  The
process $Q_t$ is equivariant, i.e., $\sp{\Psi_t} {\pov(dq) |\Psi_t}$
distributed.

Examples of Bell-type QFTs can be found in \cite{BellBeables,crea1}
and in Section~\ref{sec:example}.  It is our contention that,
essentially, there is a unique Bell-type version of every regularized
QFT.  We have to postpone, however, the discussion of operators of the
Klein--Gordon type.  We also have to assume that the QFT provides us
with the POVM $\pov(\,\cdot\,)$; this is related to an ongoing
discussion in the literature \cite{NewtonWigner,kraus,Haag}
concerning the right position operator.

\subsection{More on Identical Particles}\label{sec:identical}

The $n$-particle sector of the configuration space (without
coincidence configurations) of identical particles
$\Gommo(\RRR^3)$ is the manifold of $n$-point subsets of
$\RRR^3$; let $\conf$ be this manifold. The most common way of
describing the quantum state of $n$ fermions is by an anti-symmetric
(square-integrable) wave function $\Psi$ on $\hat\conf := \RRR^{3n}$;
let $\Hilbert$ be the space of such functions.  Whereas for bosons
$\Psi$ could be viewed as a function on $\conf$, for fermions $\Psi$
is not a function on $\conf$.

Nonetheless, the configuration observable still corresponds to a PVM
$\pov$ on $\conf$: for $B \subseteq \conf$, we set $\pov(B)
\Psi(\vq_1, \ldots, \vq_n) = \Psi(\vq_1, \ldots, \vq_n)$ if $\{\vq_1,
\ldots, \vq_n\} \in B$ and zero otherwise. In other words, $\pov(B)$
is multiplication by the indicator function of $\covering^{-1}(B)$
where $\covering$ is the obvious projection mapping $\hat\conf
\setminus \Delta \to \conf$, with $\Delta$ the set of coincidence
configurations.

To obtain other useful expressions  for this PVM, we introduce the
formal kets $|\hat{q} \rangle$ for $\hat{q} \in \hat\conf$ (to be
treated like elements of $L^2(\hat\conf)$), the anti-symmetrization
operator $S$ (i.e., the projection $L^2(\hat\conf) \to \Hilbert$), the
normalized anti-symmetrizer\footnote{The name means this: since $S$ is
a projection, $S \Psi$ is usually not a unit vector when $\Psi$ is.
Whenever $\Psi \in L^2(\hat\conf)$ is supported by a fundamental
domain of the permutation group, i.e., by a set $\Omega \subseteq
\hat\conf$ on which (the restriction of) $\covering$  is a bijection
to $\conf$, the norm of $S\Psi$ is $1/\sqrt{n!}$, so that $s\Psi$ is
again a unit vector.} $s= \sqrt{n!} \, S$, and the formal kets $|s
\hat{q}\rangle := s|\hat{q} \rangle$ (to be treated like elements of
$\Hilbert$). The $|\hat{q} \rangle$ and $|s\hat{q} \rangle$ are
normalized in the sense that
\[
   \sp{\hat{q}} {\hat{q}'} = \delta(\hat{q} - \hat{q}') \text{ and }
   \sp{s\hat{q}} {s\hat{q}'} = (-1)^{\permutation(\hat{q},\hat{q}')} \,
   \delta(q-q'),
\]
where $q=\covering(\hat{q})$, $q'=\covering(\hat{q}')$,
$\permutation(\hat{q},\hat{q}')$ is the permutation that carries
$\hat{q}$ into $\hat{q}'$ given that $q=q'$, and $(-1)^\permutation$
is the sign of the permutation $\permutation$. Now we can write
\begin{equation}\label{idenpovm}
   \pov(dq) = \sum_{\hat{q} \in \covering^{-1}(q)} |\hat{q} \rangle
   \langle \hat{q}| \, dq = n! \, S |\hat{q} \rangle \langle
   \hat{q}| \, dq = |s\hat{q} \rangle \langle s\hat{q}| \, dq,
\end{equation}
where the sum is over the $n!$ ways   of numbering the $n$
points in $q$; the last two terms actually do not depend on the choice
of $\hat{q} \in \covering^{-1}(q)$, the numbering of $q$.

The probability distribution arising from this PVM is
\begin{equation}\label{idenmeasure}
   \measure(dq) = \sum_{\hat{q} \in \covering^{-1}(q)}
   |\Psi(\hat{q})|^2 \, dq = n! \, |\Psi(\hat{q})|^2 \, dq =
   |\sp{s\hat{q}}{\Psi}|^2 \, dq
\end{equation}
with arbitrary $\hat{q} \in \covering^{-1}(q)$.

There is a way of viewing fermion wave functions as being defined on
$\conf$, rather than $\RRR^{3n}$, by regarding them as cross-sections
of a particular 1-dimensional vector bundle over $\conf$. To this end,
define an $n!$-dimensional vector bundle $E$ by
\begin{equation}\label{idenEdef}
   E_q := \bigoplus_{\hat{q} \in \covering^{-1}(q)} \CCC\,.
\end{equation}
Every function $\Psi:\RRR^{3n} \to \CCC$ naturally gives rise to a
cross-section $\Phi$ of $E$, defined by
\begin{equation}
   \Phi(q) := \bigoplus_{\hat{q} \in \covering^{-1}(q)} \Psi(\hat{q})\,.
\end{equation}
The anti-symmetric functions form a 1-dimensional subbundle of $E$
(see also \cite{identical} for a discussion of this bundle).

\section{Application to Simple Models}
\label{sec:example}

In this section, we point out how the jump rates of the model in
\cite{crea1} are contained in \eqref{tranrates} and present a
full-fledged Bell-type QFT for the second-quantized Dirac equation in
an external electromagnetic field.

Further cut-off QFTs that may provide interesting examples of
Bell-type QFTs, worth a detailed discussion in a future work
\cite{crea4}, are the scalar self-interacting field (e.g., $\Phi^4$),
QED, and other gauge field theories. We have to postpone the treatment
of these theories because they require discussions lying outside the
scope of this paper, in particular a discussion of the position
representation of photon wave functions in QED, and, concerning
$\Phi^4$, of the appropriate probability current for the Klein--Gordon
equation.

\subsection{A Simple QFT}\label{sec:crea1}

We presented a simple example of a Bell-type QFT in \cite{crea1},
and we will now briefly point to the aspects of this model that are
relevant here. The model is based on one of the simplest possible QFTs
\cite[p.~339]{Schweber}.

The relevant configuration space $\conf$ for a QFT (with a single
particle species) is the configuration space of a variable number of
identical particles in $\RRR^3$, which is the set $\Gamma(\RRR^3)$,
or, ignoring the coincidence configurations (as they are exceptions),
the set $\Gommo (\RRR^3)$ of all finite subsets of
$\RRR^3$. The $n$-particle sector of this is a manifold of dimension
$3n$; this configuration space is thus a union of (disjoint) manifolds
of different dimensions. The relevant configuration space for a theory
with several particle species is the Cartesian product of several
copies of $\Gommo (\RRR^3)$.  In the model of
\cite{crea1}, there are two particle species, a fermion and a boson,
and thus the configuration space is
\begin{equation}\label{conffermionboson}
   \conf = \Gommo (\RRR^3) \times \Gommo (\RRR^3).
\end{equation}
We will denote configurations by $q=(x,y)$ with $x$ the configuration
of the fermions and $y$ the configuration of the bosons.

For simplicity, we replaced in \cite{crea1} the sectors of $\Gommo
(\RRR^3) \times \Gommo (\RRR^3)$, which are manifolds, by vector
spaces of the same dimension (by artificially numbering the
particles), and obtained the union
\begin{equation}\label{crea1conf}
   \hat{\conf} = \bigcup_{n=0}^\infty (\RRR^3)^n \times
   \bigcup_{m=0}^\infty (\RRR^3)^m \,,
\end{equation}
with $n$ the number of fermions and $m$ the number of bosons. Here,
however, we will use \eqref{conffermionboson} as the configuration
space, since we have already discussed the space $\Gommo
(\RRR^3)$.  In comparison with \eqref{crea1conf}, this amounts to
(merely) ignoring the numbering of the particles.

$\Hilbert$ is the tensor product of a fermion Fock space and a boson
Fock space, and thus the subspace of wave functions in
$L^2(\hat{\conf})$ that are anti-symmetric in the fermion coordinates
and symmetric in the boson coordinates. Let $S$ denote the appropriate
symmetrization operator, i.e., the projection operator
$L^2(\hat{\conf}) \to \Hilbert$, and $s$ the normalized symmetrizer
\begin{equation}\label{sdef}
   s\Psi(\vx_1, \ldots, \vx_n,\vy_1, \ldots, \vy_m) = \sqrt{n!\, m!} \,
   S\Psi(\vx_1, \ldots, \vx_n,\vy_1, \ldots, \vy_m),
\end{equation}
i.e., $s = \sqrt{N! \, M!} \, S$ with $N$ and $M$ the fermion and
boson number operators, which commute with $S$ and with each other.
As in Section \ref{sec:identical}, we denote by $\covering$ the
projection mapping $\hat{\conf} \setminus \Delta \to \conf$,
$\covering(\vx_1, \ldots, \vx_n,\vy_1, \ldots, \vy_m) = (\{\vx_1,
\ldots,\vx_n\}, \{\vy_1, \ldots, \vy_m\})$.  The configuration PVM
$\pov(B)$ on $\conf$  is multiplication by
$\1_{\covering^{-1}(B)}$, which can be understood as acting on
$\Hilbert$, though it is defined on $L^2(\hat{\conf})$, since it is
permutation invariant and thus maps $\Hilbert$ to itself.  We utilize
again the formal kets $|\hat{q}\rangle$ where $\hat{q} \in \hat{\conf}
\setminus \Delta$ is a numbered configuration, for which we also write
$\hat{q} = (\hat{x},\hat{y}) = (\vx_1, \ldots, \vx_n,\vy_1, \ldots,
\vy_m)$. We also use the symmetrized and normalized kets $|s\hat{q}
\rangle = s|\hat{q} \rangle$. As   in \eqref{idenpovm}, we can write
\begin{equation}\label{crea1povm}
   \pov(dq) = \sum_{\hat{q} \in \covering^{-1}(q)} |\hat{q} \rangle
   \langle \hat{q}| \, dq = n!\, m! \, S |\hat{q} \rangle \langle
   \hat{q}| \, dq = |s\hat{q} \rangle \langle s\hat{q}| \, dq
\end{equation}
with arbitrary $\hat{q} \in \covering^{-1}(q)$. For the probability
distribution, we thus have, as  in \eqref{idenmeasure},
\begin{equation}\label{crea1measure}
   \measure(dq) = \sum_{\hat{q} \in \covering^{-1}(q)}
   |\Psi(\hat{q})|^2 \, dq  = n!\, m! \, |\Psi(\hat{q})|^2 \, dq =
   |\sp{s\hat{q}}{\Psi}|^2 \, dq
\end{equation}
with arbitrary $\hat{q} \in \covering^{-1}(q)$.

The free Hamiltonian is the second quantized Schr\"odinger operator
(with zero potential), associated with the free process described in
Section~\ref{sec:free}.  The interaction Hamiltonian is defined by
\begin{equation}\label{HIdef}
   H_\inter = \int d^3\vx \, \psi^\dag(\vx)\, (a^\dag_\profile(\vx) +
   a_{\profile}(\vx))\, \psi(\vx)
\end{equation}
with $\psi^\dag(\vx)$ the creation operators  (in position
representation), acting on the \emph{fermion} Fock space, and
$a^\dag_\profile(\vx)$ the creation operators  (in position
representation), acting on the \emph{boson} Fock space, regularized
through convolution with an $L^2$ function $\profile:\RRR^3 \to \RRR$.
$H_\inter$ has a kernel; we will now obtain a formula for it, see
\eqref{crea1kernel} below. The $|s\hat{q} \rangle$ are connected to
the creation operators according to
\begin{equation}\label{shatqpsia}
   |s\hat{q}\rangle = \psi^\dag(\vx_n) \cdots
    \psi^\dag(\vx_1) a^\dag(\vy_m) \cdots a^\dag(\vy_1) |0\rangle\,,
\end{equation}
where $|0\rangle \in \Hilbert$ denotes the vacuum state. A relevant
fact is that the creation and annihilation operators
$\psi^\dag,\psi,a^\dag$ and $a$ possess kernels. Using the canonical
(anti\nobreakdash-)commutation relations for $\psi$ and $a$, one obtains
from
\eqref{shatqpsia} the following formulas for the kernels of
$\psi(\vr)$ and $a(\vr)$, $\vr \in \RRR^3$:
\begin{align}
   \sp{s\hat{q}}{\psi(\vr)|s\hat{q}'} &= \delta_{n,n'-1} \,
   \delta_{m,m'} \,
   \delta^{3n'}(x \cup \vr -x') \, (-1)^{\permutation((\hat{x},
   \vr),\hat{x}')} \, \delta^{3m}(y-y') \label{psikernel} \\
   \sp{s\hat{q}}{a(\vr)|s\hat{q}'} &= \delta_{n,n'} \,
   \delta_{m,m'-1} \, \delta^{3n}(x-x') \,
   (-1)^{\permutation(\hat{x},\hat{x}')} \,
   \delta^{3m'}(y \cup \vr - y') \label{akernel}
\end{align}
where $(x,y) = q = \covering(\hat{q})$, and $\permutation
(\hat{x},\hat{x}')$ denotes the permutation that carries $\hat{x}$ to
$\hat{x}'$ given that $x=x'$. The corresponding formulas for
$\psi^\dag$ and $a^\dag$ can be obtained by exchanging $\hat{q}$ and
$\hat{q}'$ on the right hand sides of \eqref{psikernel} and
\eqref{akernel}.  For the smeared-out operator $a_\profile(\vr)$, we
obtain
\begin{equation}\label{aprofilekernel}
   \sp{s\hat{q}}{a_\profile(\vr)|s\hat{q}'} = \delta_{n,n'} \,
   \delta_{m,m'-1} \, \delta^{3n}(x-x') \,
   (-1)^{\permutation(\hat{x},\hat{x}')} \sum_{\vy' \in y'}
   \delta^{3m}(y- y'\setminus \vy') \, \profile(\vy' - \vr)
\end{equation}
We make use of the resolution of the identity
\begin{equation}\label{resolution}
   I = \int\limits_{\conf} dq \, |s\hat{q} \rangle \langle
   s\hat{q}|\,.
\end{equation}
Inserting \eqref{resolution} twice into \eqref{HIdef} and exploiting
\eqref{psikernel} and \eqref{aprofilekernel}, we find
\begin{equation}\label{crea1kernel}
\begin{split}
   \sp{s\hat{q}} {H_\inter| s\hat{q}'} &= \delta_{n,n'} \,
   \delta_{m-1,m'} \, \delta^{3n}(x-x') \,
   (-1)^{\permutation(\hat{x},\hat{x}')} \sum_{\vy \in y} \delta^{3m'}
   (y \setminus \vy - y') \sum_{\vx \in x} \profile(\vy - \vx) \:  \\
   &+ \delta_{n,n'} \, \delta_{m'-1,m} \,
   \delta^{3n}(x-x') \, (-1)^{\permutation(\hat{x},\hat{x}')}
   \sum_{\vy' \in y'} \delta^{3m} (y - y' \setminus \vy') \sum_{\vx \in
   x} \profile(\vy' - \vx)\,.
\end{split}
\end{equation}

By \eqref{crea1povm}, the jump rates \eqref{tranrates} are
\begin{equation}
   \sigma(q|q') = \frac{\Big[\tfrac{2}{\hbar} \, \Im \,
   \sp{\Psi}{s\hat{q}} \sp{s\hat{q}}{H_\inter| s\hat{q}'}
\sp{s\hat{q}'}{\Psi}
   \Big]^+} {\sp{\Psi}{s\hat{q}'} \sp{s\hat{q}'}{\Psi}} \,.
\end{equation}
More explicitly, we obtain from \eqref{crea1kernel} the rates
\begin{equation}\label{crea1rates}
\begin{split}
   \sigma(q|q') &= \delta_{nn'} \,\delta_{m-1,m'} \,\delta^{3n}(x-x')
   \sum_{\vy \in y} \delta^{3m'}(y\setminus \vy-y') \,
   \sigma_\crea(q'\cup \vy|q') \:  \\
   &+\delta_{nn'}\,\delta_{m,m'-1} \, \delta^{3n}(x-x') \sum_{\vy' \in
   y'} \delta^{3m}(y - y'\setminus \vy') \, \sigma_\ann(q'\setminus
   \vy'|q')
\end{split}
\end{equation}
with
\begin{subequations}
\begin{align}
   \sigma_\crea(q'\cup \vy|q')&= \frac{2 \sqrt{m'+1}}{\hbar} \,
   \frac{\Big[ \Im \, \Psi^*(\hat{q}) \,
   (-1)^{\permutation(\hat{x},\hat{x}')} \sum\limits_{\vx' \in x'}
   \varphi(\vy-\vx') \, \Psi(\hat{q}')\Big]^+}{ \Psi^*(\hat{q}') \,
   \Psi(\hat{q}')} \label{crea1crearate} \\
   \sigma_\ann(q'\setminus \vy'|q')&= \frac{2} {\hbar \sqrt{m'}}
   \,\frac{\Big[\Im \, \Psi^*(\hat{q}) \,
   (-1)^{\permutation(\hat{x},\hat{x}')} \sum\limits_{\vx' \in x'}
   \varphi(\vy'-\vx') \, \Psi(\hat{q}') \Big]^+}{ \Psi^*(\hat{q}') \,
   \Psi(\hat{q}')} , \label{crea1annrate}
\end{align}
\end{subequations}
for arbitrary $\hat{q}' \in \covering^{-1}(q')$ and $\hat{q} \in
\covering^{-1}(q)$ with $q=(x',y'\cup\vy)$ respectively $q=(x',y'
\setminus \vy')$.  (Note that a sum sign can be drawn out of the plus
function if the terms have disjoint supports.)

Equation \eqref{crea1rates} is worth looking at closely: One can read
off that the only possible jumps are $(x',y') \to (x',y' \cup \vy)$,
creation of a boson, and $(x',y') \to (x',y' \setminus \vy')$,
annihilation of a boson. In particular, while one particle is created
or annihilated, the other particles do not move. The process that we
considered in \cite{crea1} consists of pieces of Bohmian trajectories
interrupted by jumps with rates \eqref{crea1rates}; the process is
thus an example of the jump rate formula \eqref{tranrates}, and an
example of combining jumps and Bohmian motion by means of process
additivity.

The example shows how, for other QFTs, the jump rates
\eqref{tranrates} can be applied to relevant interaction Hamiltonians:
If $H_\inter$ is, in the position representation, a polynomial in the
creation and annihilation operators, then it possesses a kernel on the
relevant configuration space. A cut-off (implemented here by smearing
out the creation and annihilation operators) needs to be introduced to
make $H_\inter$ a well-defined operator on $L^2$.

If, in some QFT, the particle number operator is not conserved, jumps
between the sectors of configuration space are inevitable for an
equivariant process. And, indeed, when $H_\inter$ does not commute
with the particle number operator (as is usually the case), jumps can
occur that change the number of particles. Often, $H_\inter$ contains
\emph{only} off-diagonal terms with respect to the particle number;
then every jump will change the particle number.  This is precisely
what happens in the model of \cite{crea1}.

\subsection{Efficient Calculation of Rates in the Previous Example}
\label{sec:efficient}

We would like to give another, refined way of calculating the explicit
jump rates \eqref{crea1rates} from the definition \eqref{HIdef} of
$H_\inter$. The calculation above is rather cumbersome, partly
\emph{because} of all the $\delta$'s.  It is also striking that only
very few transitions $q' \to q$ are actually possible, which suggests
that it is unnecessary to write down a formula for the kernel
$\sp{q}{H_\inter|q'}$ valid for all pairs $q,q'$. Rather than writing
down all the $\delta$ terms as in \eqref{crea1rates}, it is easier
to specify the possible transitions $q' \to q$ and to write down the
rates, such as \eqref{crea1crearate} and \eqref{crea1annrate}, only
for these transitions. Thus, for a more efficient calculation of the
rates, it is advisable to first determine the possible transitions,
and then we need keep track only of the corresponding kernel elements.

\subsubsection{A Diagram Notation}

To formulate this more efficient strategy, it is helpful to regard 
$\Psi$
as a cross-section of a fiber bundle $E$ over the Riemannian manifold
$\conf$, or of a countable union $E = \bigcup_i E^{(i)}$ of bundles
$E^{(i)}$ over Riemannian manifolds $\conf^{(i)}$ with $\conf = 
\bigcup_i
\conf^{(i)}$. (In the present example, with $\conf$ given by
\eqref{conffermionboson},  we take $i$ to be the pair $(n,m)$ of 
particle
numbers, $\conf^{(n,m)}$ to be the $(n,m)$-particle sector, and 
$E^{(i)}$
to be defined by \eqref{idenEdef} (with $\covering$ the natural 
projection
from $\hat{\conf} \setminus \Delta$, with $\hat{\conf}$ given by
\eqref{crea1conf}, to $\conf$).  The $\hat{q} \in \covering^{-1}(q)$ can
be
viewed as defining an orthonormal basis of $E_q$.)

A key element of the strategy is a special diagram notation for
operators. The operators we have in mind are $H_\inter$ and its
building blocks, the field operators. The strategy will start with the
diagrams for the field operators, and obtain from them a diagram for
$H_\inter$. The diagram will specify, for an operator $O$, what the
kernel of $O$ is, while leaving out parts of the kernel that are zero. 
So
let us assume that $O$ has kernel $\sp{q}{O|q'}$, i.e., $(O\Psi)(q) =
\int \sp{q}{O|q'} \, \Psi(q') \, dq'$. The diagram
\begin{equation}\label{arrow}
   q' \xrightarrow[O]{K(q',\lambda)} F(q',\lambda)
\end{equation}
means that the operator $O$ has \emph{kernel constructed from $F$ and
$K$},
\begin{equation}\label{kernelarrow}
   \sp{q}{O|q'} = \int\limits_{\Lambda} d\lambda \, \delta\big(q-
   F(q',\lambda)\big) \, K(q',\lambda),
\end{equation}
where $\lambda$ varies in some parameter space $\Lambda$, $F: \conf
\times \Lambda \to \conf$, and $K$ is a function (or distribution) of
$q'$ and $\lambda$ such that
   $K(q',\lambda) : E_{q'} \to E_{F(q',\lambda)}$
is a $\CCC$-linear mapping.

The role of $\lambda$ is to parametrize the possible transitions;
e.g., for the boson creation \eqref{crea1crearate} in the previous
section, $\lambda$ would be the position $\vy$ of the new boson, and
$\Lambda = \RRR^3$.  The notation \eqref{arrow} does not explicitly
mention what $\Lambda$ and the measure $d\lambda$ are; this will
usually be clear from the context of the diagram.  The measure
$d\lambda$ will usually be a uniform distribution over the parameter
space $\Lambda$, such as Lebesgue measure if $\Lambda = \RRR^d$ or the
counting measure if $\Lambda$ is finite or countably infinite. We may
also allow having a different $\Lambda_{q'}$ for every $q'$.

In words, \eqref{arrow} may be read as: ``According to $O$, the
possible transitions from $q'$ are to $F(q',\lambda)$, and are
associated with the amplitudes $K(q',\lambda)$.'' In fact, when $O =
H$, a jump from $q'$ can lead only to those $q$'s for which $q =
F(q',\lambda)$ for some value of $\lambda$, and the corresponding jump
rate \eqref{tranrates} is
\begin{equation}\label{arrowrates}
   \sigma\big(F(q',\lambda)\big|q'\big) = \frac{[(2/\hbar) \, \Im \,
   \Psi^*(F(q',\lambda)) \, K(q',\lambda) \, \Psi(q')]^+} {\Psi^*(q')
   \, \Psi(q')},
\end{equation}
provided that for given $q'$, $F(q', \,\cdot\,)$ is an injective
mapping.  Here, $\sigma(q|q')$ is the density of the measure
$\sigma(dq|q')$ with respect to the measure on $\conf$
\begin{equation}\label{arrowuniform}
   \mu_{q'}(dq) = \int\limits_{\Lambda} d\lambda \, \delta\big(
   q-F(q',\lambda) \big) \, dq,
\end{equation}
where $\delta(q-q_0) \, dq$ denotes the measure on $\conf$ with total
weight 1 concentrated at $q_0$. \eqref{arrowuniform}, the image of
$d\lambda$ under the map $F(q',\cdot\,)$,  is concentrated on the set 
$\{
F(q',\lambda) : \lambda \in \Lambda\}$ of possible destinations and 
plays
the role of the ``uniform distribution''  over this set. In other words,
\eqref{arrowrates} is the rate of occurrence, with respect to 
$d\lambda$,
of the transition corresponding to $\lambda$.  (For the boson creation
rate \eqref{crea1crearate}, $\mu_{q'}(dq)$ turns out the Lebesgue 
measure
in $\vy$ on the subset $\{q' \cup \vy: \vy \in \RRR^3 \setminus q'\}
\subseteq \conf$.)

Given $O$, the choice of $\Lambda, F$, and $K$ is not unique. One
could always choose $\Lambda = \conf$, $F(q',q) = q$, and $K(q',q) =
\sp{q}{O|q'}$, which of course would mean to miss the point of this
notation.  The case that $F$ and $K$ do not depend on a parameter
$\lambda$ is formally contained in the scheme \eqref{kernelarrow} by
taking $\Lambda$ to be a one-point set (and $d\lambda$ the counting
measure); in this case \eqref{kernelarrow} means
\begin{equation}\label{kernelarrownolambda}
   \sp{q}{O|q'} = \delta(q-F(q')) \, K(q')\,.
\end{equation}
Conversely, whenever $\# \Lambda =1$, the dependence of $F$ and $K$ on
the parameter $\lambda$ is irrelevant.

A basic advantage of the notation \eqref{arrow}, compared to writing
down a formula for $\sp{q}{O|q'}$, is that many $\delta$ factors
become unnecessary.  For example,  if $O$ is  multiplication  by
$V(q)$, then ($\Lambda$ is a one-point set and) we have the diagram
\[
   q' \xrightarrow[O]{V(q')} q'.
\]

\subsubsection{Operations With Diagrams}

For the product $O_2O_1$ of two operators given by diagrams, we have
the diagram
\begin{equation}\label{productarrow}
   q' \xrightarrow[O_2O_1]{K_2(F_1(q',\lambda_1),\lambda_2) \,
   K_1(q',\lambda_1)} F_2(F_1(q',\lambda_1),\lambda_2)
\end{equation}
with parameter space $\Lambda_1 \times \Lambda_2$, for which we
also write
\begin{equation}\label{concatarrow}
   q' \xrightarrow[O_1]{K_1(q',\lambda_1)} F_1(q',\lambda_1)
   \xrightarrow[O_2]{K_2(F_1(q',\lambda_1),\lambda_2)}
   F_2(F_1(q',\lambda_1),\lambda_2).
\end{equation}
We  thus define  the concatenation of two diagrams by means of
the composition of the transition mappings and the product of the
amplitudes,
i.e., using obvious notation,
\begin{equation}\label{shortconcatarrow}
   q_1 \xrightarrow{\alpha} q_2 \xrightarrow{\beta} q_3 \quad
   \text{means} \quad q_1 \xrightarrow{\alpha\beta} q_3.
\end{equation}
Thus, multiplication of operators corresponds to concatenation of
diagrams.

For the sum $O_1 +O_2$ of two operators given by diagrams with the
same parameter space $\Lambda_1 = \Lambda _2 = \Lambda$ and the same
transition mapping $F_1(q', \lambda) = F_2(q',\lambda) =
F(q',\lambda)$, we have the diagram
\begin{equation}\label{sumarrow}
   q' \xrightarrow[O_1 + O_2]{K_1(q',\lambda) + K_2(q',\lambda)}
   F(q',\lambda).
\end{equation}

\subsubsection{Diagrams of Creation and Annihilation Operators}

We now write down diagrams for creation and annihilation operators.
In the case that $O = O(\vr)$ arises from formally evaluating an
operator-valued distribution $O(\vx)$ at $\vx = \vr$, the dependence
of $K(q',\lambda)$ on $\lambda$ is  in the sense of distributions
rather than functions. More precisely, we have
\begin{equation}\label{Kdistribution}
   K(q',\lambda) = D(q',\lambda) \, K_0(q',\lambda)
\end{equation}
where $D$ is a (real-valued) distribution on $\conf \times \Lambda$,
and $K_0$ a mapping-valued function such that for every $q'$ and
$\lambda$, $K_0(q',\lambda)$ is a linear mapping $E_{q'} \to
E_{F(q',\lambda)}$.

For $\psi^\dag(\vr)$ and $\psi(\vr)$, $\vr \in \RRR^3$, we have
(recall that $x'$ is a finite subset of $\RRR^3$)
\begin{subequations}
\begin{align}
   (x',y')& \xrightarrow[\psi^\dag(\vr)] {\alpha_\fer} (x' \cup \vr,y')
   \qquad\quad (\#\Lambda =1) \\ (x',y')& \xrightarrow[\psi(\vr)]
   {\delta(\vx'-\vr) \, \varepsilon_\fer} (x' \setminus \vx',y') \qquad
   (\Lambda = x', \lambda=\vx')
\end{align}
\end{subequations}
using linear mappings $\alpha_\fer: E_{q'} \to E_{ (x' \cup \vr,y')}$
(``append a fermion'') and $\varepsilon_\fer: E_{q'} \to E_{ (x'
\setminus \vx',y')}$ (``erase a fermion''), which can be regarded as
the natural mappings between these fiber spaces. They are defined
through the following properties:
\begin{subequations}
\begin{align}
   &\alpha_\fer \Psi\text{ is appropriately symmetrized} \\
   &\big(\alpha_\fer \Psi\big)((\hat{x}', \vr), \hat{y}') =
\frac{1}{\sqrt{n'
   +1}} \, \Psi(\hat{x}',\hat{y}') \\
   &\big(\varepsilon_\fer \Psi\big) (\hat{x},\hat{y}') =
   \sqrt{n'} \, \Psi((\hat{x}, \vx'),\hat{y}')
\end{align}
\end{subequations}
where $\Psi \in E_{q'}$, and $\hat{x}$ is an arbitrary ordering of the
set $x=x' \setminus \vx'$. (Recall that the set $\covering^{-1}(q')$ of
the possible orderings of $q'$ forms a basis of $E_{q'}$, so that
every ordering $(\hat{x}', \hat{y}') = \hat{q}' \in
\covering^{-1}(q')$ corresponds to a particular component of
$\Psi$. Thus, $((\hat{x}',\vr),\hat{y}') \in \covering^{-1}(x' \cup \vr,
y')$ corresponds to a particular component in $E_{(x' \cup \vr, y')}$.)

For the smeared-out creation and annihilation operators
$a_\profile^\dag(\vr)$ and $a_\profile(\vr)$, we have
\begin{subequations}
\begin{align}
   (x',y') &\xrightarrow[a_\profile^\dag(\vr)] {\profile(\vy-\vr) \,
   \alpha_\bos} (x',y' \cup \vy) \qquad (\Lambda = \RRR^3, \lambda =
   \vy) \\
   (x',y') &\xrightarrow[a_\profile(\vr)] {\profile(\vy'-\vr) \,
   \varepsilon_\bos} (x', y' \setminus \vy') \qquad (\Lambda = y',
   \lambda = \vy')
\end{align}
\end{subequations}
where $\alpha_\bos$ (``append a boson'') and $\varepsilon_\bos$
(``erase a boson'') are the analogous linear mappings relating
different spaces, $\alpha_\bos: E_{q'} \to E_{(x',y' \cup \vy)}$ and
$\varepsilon_\bos: E_{q'} \to E_{(x',y' \setminus \vy')}$, defined by
the following properties:
\begin{subequations}
\begin{align}
   &\alpha_\bos \Psi\text{ is appropriately symmetrized} \\
   &\big(\alpha_\bos \Psi\big) (\hat{x}',(\hat{y}', \vy)) =
   \frac{1}{\sqrt{m'+1}} \, \Psi(\hat{x}',\hat{y}') \\
   &\big(\varepsilon_\bos \Psi\big) (\hat{x}', \hat{y}) =
   \sqrt{m'} \, \Psi(\hat{x}', (\hat{y}, \vy')),
\end{align}
\end{subequations}
where $\hat{y}$ is an arbitrary ordering of the set $y=y' \setminus
\vy'$, $\hat{x}'$ one of $x'$, $\hat{y}'$ one of $y'$, and $\Psi \in
E_{q'}$.

\subsubsection{Application of the Diagram Method}

Now let us apply the strategy to the example \eqref{HIdef} of the
previous section. For $\psi^\dag(\vr) \, a^\dag_\profile(\vr) \,
\psi(\vr)$, we have the diagram
\[
   q' \xrightarrow[\psi(\vr)]{\delta(\vx'-\vr) \, \varepsilon_\fer} (x'
   \setminus \vx', y') \xrightarrow[a^\dag_\profile(\vr)]{\profile(\vy
   -\vr) \, \alpha_\bos} (x' \setminus \vx',y' \cup \vy)
   \xrightarrow[\psi^\dag(\vr)] {\alpha_\fer} (x' \setminus \vx' \cup
   \vr,y' \cup \vy)
\]
with $\Lambda = x' \times \RRR^3$. Using the concatenation rule
\eqref{shortconcatarrow}, we can write instead
\[
   q' \xrightarrow[\psi^\dag(\vr) \, a^\dag_\profile(\vr) \,
   \psi(\vr)]{\delta(\vx'-\vr) \,\profile(\vy -\vr) \, \alpha_\fer
   \alpha_\bos \varepsilon_\fer} (x' \setminus \vx' \cup \vr,y' \cup 
\vy).
\]
Integrating over $d\vr$, we obtain, since $x' \setminus \vx' \cup \vr$ 
may
be replaced by $x'$, which is independent of $\vx'$,
\begin{equation}\label{creaarrow}
   q' \xrightarrow[\int d\vr \, \psi^\dag(\vr) \, a^\dag_\profile(\vr)
   \, \psi(\vr)]{\sum\limits_{\vx' \in x'}\profile(\vy -\vx') \,
   \alpha_\fer \alpha_\bos \varepsilon_\fer} (x',y' \cup \vy),
\end{equation}
with $\Lambda =  \RRR^3$.  We have now taken care of one of two terms in
\eqref{HIdef}, involving
$a^\dag$ rather than $a$. {}From \eqref{creaarrow} we read off, without
a big calculation, that this term corresponds to jumps $(x',y') \to
(x',y'\cup \vy)$, or creation of a boson. The corresponding jump rate
is given by \eqref{arrowrates}, and reads here:
\begin{equation}\label{crea1arrowcrearate}
   \sigma(x',y' \cup \vy|q') = \frac{2}{\hbar} \, \frac{\Big[\Im \,
   \Psi^*(x',y' \cup \vy) \sum\limits_{\vx' \in x'} \profile(\vy -\vx')
   \, \alpha_\fer \alpha_\bos \varepsilon_\fer \, \Psi(q')\Big]^+}
   {\Psi^*(q') \, \Psi(q')}.
\end{equation}
This result agrees with \eqref{crea1crearate}.\footnote{Here is why:
First, $\Psi^*(q') \, \Psi(q') = n'! \, m'! \, \Psi^*(\hat{q}') \,
\Psi(\hat{q}')$ because the inner product in $E_{q'}$ involves
summation over all $\hat{q}' \in \covering^{-1}(q')$. Similarly, the
square bracket in the numerator of \eqref{crea1arrowcrearate}
involves the inner product of $E_{(x',y' \cup \vy')}$, consisting of
$n'! \, (m'+1)!$ contributions. The numberings $\hat{q}$ and
$\hat{q}'$ in \eqref{crea1crearate} can be so chosen that $\hat{x} =
\hat{x}'$, $\vx'$ gets the last place of $\hat{x}'$, and $\hat{y} =
\hat{y}' \cup \vy'$; then $\permutation(\hat{x},\hat{x}')$ is trivial,
and $\alpha_\fer \alpha_\bos \varepsilon_\fer \Psi(\hat{q}) = 
(n')^{-1/2}
(m'+1)^{-1/2} (n')^{1/2} \Psi(\hat{q}')$. Thus, the square bracket in
\eqref{crea1arrowcrearate} is $n'! \, m'!\sqrt{m'+1}$ times the
square bracket in \eqref{crea1crearate}.}

We treat the term $\int d\vr \, \psi^\dag(\vr) \, a_\profile(\vr) \,
\psi(\vr)$ in the same way: We begin with the diagram
\[
   q' \xrightarrow[\psi(\vr)]{\delta(\vx'-\vr) \, \varepsilon_\fer} (x'
   \setminus \vx', y') \xrightarrow[a_\profile(\vr)]{\profile(\vy'
   -\vr) \, \varepsilon_\bos} (x' \setminus \vx',y' \setminus \vy')
   \xrightarrow[\psi^\dag(\vr)] {\alpha_\fer} (x' \setminus \vx' \cup
   \vr,y' \setminus \vy')
\]
with $\Lambda = x' \times y'$. Then we integrate over $d\vr$ and
obtain the associated jump rate
\begin{equation}\label{crea1arrowannrate}
   \sigma(x',y' \setminus \vy'|q') = \frac{2}{\hbar} \, \frac{\Big[\Im
   \, \Psi^*(x',y' \setminus \vy') \sum\limits_{\vx' \in x'}
   \profile(\vy' -\vx') \, \alpha_\fer \varepsilon_\bos \varepsilon_\fer 
\,
   \Psi(q')\Big]^+} {\Psi^*(q') \, \Psi(q')},
\end{equation}
which agrees with \eqref{crea1annrate}.  Finally, $H_\inter$ (the sum of
both contributions) corresponds according to \eqref{tranrates} to jumps
which, since the two contributions have no transitions $q' \to q$ in
common
(or, in other words, since their kernels have disjoint supports in 
$\conf
\times \conf$), are \emph{either} $q' \to (x',y' \cup \vy)$, with rate
\eqref{crea1arrowcrearate}, \emph{or} $q' \to (x',y' \setminus \vy')$,
with
rate \eqref{crea1arrowannrate}.

\subsection{Pair Creation in an External Field}
\label{sec:positron}

As our second example, we present the Bell-type version of a
reasonable and often used QFT of electrons and positrons, in which the
electromagnetic field is a background field \cite{Ruijsenaars}.  The
Bell-type version exhibits pair creation and annihilation (in the
literal sense) and employs various notions we have introduced: process
additivity, the configuration space $\Gommo(\RRR^3)$ of a variable
number of identical particles, the free process, POVMs which are not
PVMs, and stochastic jumps.

\subsubsection{Fock Space and Hamiltonian}\label{sec:posiFock}

We consider the second quantized Dirac field in an electromagnetic
background field $A_\mu(\vx,t)$. In terms of field operators, the
Hamiltonian reads
\begin{equation}\label{fieldhamil}
   H= \int d^3 x :{\Phi^*}(\vx)\big[-\I c\hbar \boldsymbol{\alpha} \cdot
   \nabla +\beta m c^2+ e(\boldsymbol{\alpha}\cdot\boldsymbol{A} +
   A_0) \big]\Phi(\vx):\;\;,
\end{equation}
with colons denoting normal ordering. Note that $H$ is time-dependent
due to the time-dependence of $A_\mu(\vx,t)$; more precisely,
$H_\inter$ is time-dependent while $H_0$ is fixed. As a consequence,
the relevant jump rate \eqref{tranrates} is now time-dependent in
three ways: through $H_\inter$, through $\Psi$, and through $q' =
Q_t$.

We quickly recall what the Hilbert space and the field operators are,
and specify what POVM we use.  After that, we construct the associated
process.

The Hilbert space $L^2(\RRR^3,\CCC^4)$ of the Dirac equation is split
into the orthogonal sum $\Hilbert_+ \oplus \Hilbert_-$ of the positive
and negative energy subspaces of the \emph{free} Dirac operator,
\[
   h_0= -\I c\hbar \valpha \cdot \nabla + \beta mc^2\,.
\]
The 1-electron Hilbert space $\Hilbert_\el$ and the 1-positron Hilbert
space $\Hilbert_\pos$ are copies of $\Hilbert_+$, and the Fock space
$\Fock=\Gamma \Hilbert^{(1)}$ arises then from the one-particle
Hilbert space $\Hilbert^{(1)} = \Hilbert_\el \oplus \Hilbert_\pos$ in
the usual manner: with the anti-symmetrization operator $\Anti$,
\begin{equation}\label{elplusposFock}
   \Fock= \bigoplus_{N=0}^\infty \Anti
   (({\Hilbert_\el}\oplus{\Hilbert_\pos})^{\otimes N})\,,
\end{equation}
which can be naturally identified with
\begin{equation}\label{elFockposFock}
   \Hilbert := \Fock_\el \otimes \Fock_\pos =
   \bigoplus_{n=0}^\infty \Anti (\Hilbert_\el^{\otimes n}) \otimes
   \bigoplus_{\pn=0}^\infty \Anti (\Hilbert_\pos^{\otimes \pn})\,.
\end{equation}
Since $\Hilbert_+ \subseteq L^2(\RRR^3,\CCC^4)$, $\Hilbert$ can be
understood as a subspace of
\begin{equation}\label{elHext}
   \Hilbert_\ext := \bigoplus_{n=0}^\infty
   \Anti(L^2(\RRR^3,\CCC^4)^{\otimes n}) \otimes
   \bigoplus_{\pn=0}^\infty \Anti(L^2(\RRR^3,\CCC^4)^{\otimes \pn}) .
\end{equation}

  We choose the POVM and configuration space in the way suggested by the
form \eqref{elFockposFock}, rather than \eqref{elplusposFock}:
\begin{equation}
   \conf = \Gommo(\RRR^3) \times \Gommo(\RRR^3),
\end{equation}
where the first factor represents electrons and the second
positrons. (Recall from Section \ref{sec:free} that
$\Gommo(\RRR^3)$ denotes the space of all finite subsets
of $\RRR^3$.  Another interesting possibility, suggested by the
representation (\ref{elplusposFock}), is to set $\conf =
\Gommo(\RRR^3)$. This would mean that, insofar as the
configuration is concerned, electrons and positrons are not
distinguished. However, we will not pursue this possibility here.)
The natural POVM $\pov$ (see Section~\ref{sec:GammaPOVM} and
Section~\ref{sec:identical}) can be expressed as an extension from
rectangular sets (the existence of such an extension is proved in
Section~4.4 of \cite{crea2A}):
\[
   \pov(B_\el \times B_\pos) = \Gamma\pov^{(1)}(B_\el) \otimes
   \Gamma\pov^{(1)}(B_\pos)
\]
with $\pov^{(1)}$ the POVM on $\Hilbert_+$ that we considered before,
arising by projection from the natural PVM on $L^2(\RRR^3,\CCC^4)$.
Alternatively, $\pov$ can be viewed as arising, by projection to
$\Hilbert$, and from $\hat{\conf} = \bigcup_{n=0}^\infty (\RRR^3)^n
\times \bigcup_{\pn=0}^\infty (\RRR^3)^\pn$ to $\conf$, of the natural
PVM on $\hat{\conf}$ acting on $\Hilbert_\ext$.  Note that $\pov$
represents the usual $|\Psi|^2$ distribution in the sense that for a
configuration $q$ with electrons at $\vx_1, \ldots, \vx_n$ and
positrons at $\pvx_1, \ldots, \pvx_\pn$, we have
\[
   \measure(dq) = \sp{\Psi}{\pov(dq)|\Psi} = n! \pn! \,
   |\Psi^{(n,\pn)}(\vx_1, \ldots, \pvx_\pn)|^2 \, d\vx_1 \cdots d\pvx_\pn
\]
where $\Psi^{(n,\pn)}$ is just the wave function $(\RRR^3)^{n+\pn}\to
(\CCC^4)^{\otimes (n+\pn)}$ we get when we decompose the state vector
in the manner suggested by \eqref{elHext}. $\Psi$ is normalized so
that
\[
   \sum_{n,\pn =0}^\infty \int d\vx_1 \cdots d\pvx_\pn \,
   |\Psi^{(n,\pn)}(\vx_1, \ldots, \pvx_\pn)|^2 = 1.
\]

The field operator is defined by
\begin{equation}\label{Phidef}
   \Phi(f) = b(P_+ f) + d^*(CP_- f)
\end{equation}
where $f$ is a test function from $L^2(\RRR^3,\CCC^4)$, $P_\pm$ is the
projection to $\Hilbert_\pm \subseteq L^2(\RRR^3,\CCC^4)$, $C$ is the
charge conjugation operator which maps $\Hilbert_-$ to $\Hilbert_+$
and vice versa, and $b$ is the electron annihilation and $d^*$ the
positron creation operator. Letting $\ve_\Dindex$ be the standard
orthonormal basis of $\CCC^4$, $\Dindex =1,2,3,4$, $\Phi(\vx)$ stands
for $\Phi_i(\vx) = \Phi(\ve_\Dindex \, \delta(\,\cdot\, -\vx))$, where
$\Dindex$ gets contracted with the $\valpha$ matrices.  Similarly, we
define, as usual,
\begin{subequations}
\begin{align}
   b_\Dindex(\vx) &= b\Big(P_+(\ve_\Dindex \,
   \delta(\,\cdot\, - \vx)) \Big) \\
   \text{and } d_\Dindex(\vx) &= d \Big(CP_-(\ve_\Dindex \,
   \delta(\,\cdot\, - \vx)) \Big).
\end{align}
\end{subequations}
We thus have $\Phi_\Dindex(\vx) =
b_\Dindex(\vx) + d_\Dindex^*(\vx)$.

\subsubsection{The Associated Process}

We now describe the associated Markov process. The free part of
(\ref{fieldhamil}),
\[
   H_0= \int d^3 x :{\Phi^*}(\vx)\big[-\I c \hbar
   \valpha\cdot\nabla +\beta m c^2 \big]\Phi(\vx):\;\;,
\]
preserves particle numbers (it commutes with the electron and
positron number operators), evolving the $(n,\pn)$-particle sector
of the Fock space according to the free $(n,\pn)$-particle Hamiltonian
\[
   H^{(n,\pn)}_0 = \sum_{k=1}^n h^{(k)}_0 + \sum_{\pk=1}^\pn
   \ph^{(\pk)}_0\,,
\]
with
\begin{align}
   h^{(k)}_0 &= -\I c\hbar \valpha^{(k)} \cdot \nabla_k + \beta^{(k)} 
mc^2
   \nonumber\\
   \ph^{(\pk)}_0 &= -\I c\hbar \palpha^{(\pk)} \cdot 
\widetilde{\nabla}_\pk
   + \widetilde\beta^{(\pk)} mc^2\,,\nonumber
\end{align}
where $\valpha^{(k)}$ and $\beta^{(k)}$ act on the $k$-th electron
index in the tensor product representation \eqref{elFockposFock} and
$\palpha^{(\pk)}$ and $\widetilde{\beta}^{(\pk)}$ on the $\pk$-th
positron index. $\widetilde{\nabla}_\pk$ is the gradient with
respect to $\pvx_\pk$.

With $H_0$ is associated a deterministic motion of the configuration
in $\conf$, the free process introduced in Section \ref{sec:free}.
During this motion, the actual numbers $N, \pN$ of electrons and
positrons remain constant, while the positions $(\vX_1, \ldots, \vX_N,
\pvX_1, \ldots, \pvX_\pN)=:Q$ move according to Bohm--Dirac velocities
(\ref{BohmDirac}), i.e.
\begin{subequations}\label{elposmotion}
\begin{align}
   \dot{\vX}_k &= c\frac{\Psi^*(Q) \, \valpha^{(k)} \, \Psi(Q)}
   {\Psi^*(Q) \, \Psi(Q)} \\
   \dot{\pvX}_\pk &= c\frac{\Psi^*(Q) \, \palpha^{(\pk)} \, \Psi(Q)}
   {\Psi^*(Q) \, \Psi(Q)}
\end{align}
\end{subequations}
where numerators and denominators are scalar products in
$(\CCC^4)^{\otimes (N+\pN)}$.

We turn now to the interaction part. Setting $A =\valpha \cdot
e\boldsymbol{A} + e A_0$, we have that
\begin{subequations}
\begin{align}
   H_\inter&= \int d^3 \vx :{\Phi^*}(\vx) \, A(\vx)\,\Phi(\vx):\;\; =\\
   &= \sum_{\Dindex,\Dindextwo=1}^4 \int d^3 \vx :(b^*_\Dindex(\vx) +
   d_\Dindex(\vx)) \, A^{\Dindex,\Dindextwo} (\vx) \,
   (b_\Dindextwo(\vx)+d^*_\Dindextwo(\vx)):\;\; = \\
\begin{split} \label{HIterms}
   &= \sum_{\Dindex,\Dindextwo=1}^4 \int d^3 \vx \, \Big(b^*_\Dindex(\vx)
\,
   A^{\Dindex,\Dindextwo}(\vx) \, b_\Dindextwo(\vx) + d_\Dindex(\vx)
   \,A^{\Dindex,\Dindextwo}(\vx) \, b_\Dindextwo(\vx) \: + \\
   &\quad + \: b^*_\Dindex(\vx) \, A^{\Dindex,\Dindextwo}(\vx) \,
   d^*_\Dindextwo(\vx) - d^*_\Dindextwo(\vx) \,
   A^{\Dindex,\Dindextwo}(\vx) \, d_\Dindex(\vx) \Big).
\end{split}
\end{align}
\end{subequations}
Since $H_\inter$ is a polynomial in creation and annihilation
operators, it possesses a kernel and corresponds to stochastic
jumps. To compute the rates, we apply the strategy developed in
Section \ref{sec:efficient}, using diagrams. To this end, we regard
fermionic wave functions again as cross-sections of a bundle $E$,
defined here by
\begin{equation}\label{elposEdef}
   E_q = \bigoplus_{\hat{q} \in \covering^{-1}(q)} (\CCC^4)^{\otimes n}
   \otimes (\CCC^4)^{\otimes \pn}.
\end{equation}
Fermionic symmetry of a cross-section $\Psi$ of $E$ means that
\begin{equation}\label{Psiantisym}
   \Psi\!\!
   \begin{array}{l}
     {\scriptstyle \permutation(\Dindex_1 \ldots \Dindex_n),
     \ppermutation(\pDindex_1 \ldots \pDindex_\pn)} \\
     (\permutation(\vx_1 \ldots \vx_n), \ppermutation(\pvx_1 \ldots
     \pvx_\pn)) \\ {}
   \end{array}
   = (-1)^\permutation \, (-1)^\ppermutation \, \Psi\!\!
   \begin{array}{l}
     {\scriptstyle \Dindex_1 \ldots \Dindex_n, \pDindex_1 \ldots
     \pDindex_\pn} \\
     (\vx_1 \ldots \vx_n, \pvx_1 \ldots \pvx_\pn) \\ {}
   \end{array}
\end{equation}
for all permutations $\permutation \in S_n$ and $\ppermutation \in
S_{\pn}$.

The diagrams for $b^*_\Dindex(\vx),b_\Dindex(\vx),d^*_\Dindex(\vx)$,
and $d_\Dindex(\vx)$ are
\begin{subequations}
\begin{align}
   (x',\px') &\xrightarrow [b^*_\Dindex(\vx)]
   {\sum_\Dindextwo {S_+}^{\Dindextwo}_{\Dindex} (\vx' - \vx) \,
   \alpha_\el(\ve_\Dindextwo)} (x'\cup \vx',\px')\\
   (x',\px') &\xrightarrow [b_\Dindex(\vx)]
   {\sum_\Dindextwo {S_+}^{\Dindextwo}_{\Dindex} (\vx' - \vx) \,
   \varepsilon_\el(\ve_\Dindextwo)} (x'\setminus \vx',\px')\\
   (x',\px') &\xrightarrow [d^*_\Dindex(\vx)]
   {\sum_\Dindextwo {S_-}^{\Dindextwo}_{\Dindex} (\pvx' - \vx) \,
   \alpha_\pos(\ve_\Dindextwo)} (x',\px'\cup \pvx')\\
   (x',\px') &\xrightarrow [d_\Dindex(\vx)]
   {\sum_\Dindextwo {S_-}^{\Dindextwo}_{\Dindex} (\pvx' - \vx) \,
   \varepsilon_\pos(\ve_\Dindextwo)} (x',\px'\setminus \pvx')
\end{align}
\end{subequations}
where the matrix function ${S_+}_\Dindextwo^\Dindex (\vx)$ is defined
as the $\Dindextwo$-component of $P_+ (\ve_\Dindex \,
\delta(\,\cdot\,))$, and ${S_-}_\Dindextwo^\Dindex (\vx)$ as the
$\Dindextwo$-component of $CP_- (\ve_\Dindex \, \delta(\,\cdot\,))$.
The linear mappings $\alpha_\el(\ve_\Dindextwo): E_{q'} \to E_{(x'
\cup \vx',\px')}$ (``append an electron with spinor
$\ve_\Dindextwo$'') and $\varepsilon_\el(\ve_\Dindextwo): E_{q'} \to
E_{(x' \setminus \vx',\px')}$ (``erase an electron, contracting with
spinor $\ve_\Dindextwo$'') are defined through their properties that
for $\Psi \in E_{q'}$,
\begin{subequations}
\begin{align}
   &\alpha_\el \Psi \text{ is appropriately symmetrized} \\
   &\big(\alpha_\el(\ve_\Dindextwo) \Psi\big) ((\hat{x}', 
\vx'),\hat{\px'})
=
   \frac{1}{\sqrt{n'+1}} \, \Psi(\hat{x}',\hat{\px'}) \otimes
   \ve_\Dindextwo \\
   &\big(\varepsilon_\el(\ve_\Dindextwo) \Psi\big) (\hat{x}, \hat{\px'}) 
=
   \sqrt{n'} \, \Psi_\Dindextwo ((\hat{x}, \vx'),\hat{\px'}),
\end{align}
\end{subequations}
where $\hat{x}$ is an arbitrary  ordering of $x=x' \setminus \vx'$,
$\hat{x}'$ one of $x'$, and $\hat{\px'}$ one of $\px'$. We refer to
the last electron slot when writing the tensor product or taking the
$\Dindextwo$-component.  $\alpha_\pos (\ve_\Dindextwo)$ and
$\varepsilon_\pos(\ve_\Dindextwo)$ are defined analogously.

For the four terms in \eqref{HIterms}, we thus get the four diagrams
(omitting the multiplication by $A^{\Dindex,\Dindextwo}(\vx)$)
\begin{subequations}
\begin{align}
   (x',\px') &\xrightarrow [b_\Dindextwo(\vx)] {\sum_\Dindexthree
   {S_+}^{\Dindexthree}_{\Dindextwo} (\vx' - \vx) \,
   \varepsilon_\el(\ve_\Dindexthree)} (x'\setminus \vx',\px') 
\xrightarrow
   [b^*_\Dindex(\vx)] {\sum_\Dindexfour {S_+}^{\Dindexfour}_{\Dindex}
   (\vx'' - \vx) \, \alpha_\el(\ve_\Dindexfour)} (x'\setminus \vx' \cup
   \vx'',\px') \\
   (x',\px') &\xrightarrow [b_\Dindextwo(\vx)] {\sum_\Dindexthree
   {S_+}^{\Dindexthree}_{\Dindextwo} (\vx' - \vx) \,
   \varepsilon_\el(\ve_\Dindexthree)} (x'\setminus \vx',\px') 
\xrightarrow
   [d_\Dindex(\vx)] {\sum_\Dindexfour {S_-}^{\Dindexfour}_{\Dindex}
   (\pvx' - \vx) \, \varepsilon_\pos(\ve_\Dindexfour)} (x'\setminus
   \vx',\px'\setminus \pvx')\\
   (x',\px') &\xrightarrow [d^*_\Dindextwo(\vx)] {\sum_\Dindexthree
   {S_-}^{\Dindexthree}_{\Dindextwo} (\pvx' - \vx) \,
   \alpha_\pos(\ve_\Dindexthree)} (x',\px'\cup \pvx') \xrightarrow
   [b^*_\Dindex(\vx)] {\sum_\Dindexfour {S_+}^{\Dindexfour}_{\Dindex}
   (\vx' - \vx) \, \alpha_\el(\ve_\Dindexfour)} (x'\cup \vx',\px' \cup
   \pvx')\\
   (x',\px') &\xrightarrow [d_\Dindex(\vx)] {\sum_\Dindexthree
   {S_-}^{\Dindexthree}_{\Dindex} (\pvx' - \vx) \,
   \varepsilon_\pos(\ve_\Dindexthree)} (x',\px'\setminus \pvx')
\xrightarrow
   [d^*_\Dindextwo(\vx)] {\sum_\Dindexfour
   {S_-}^{\Dindexfour}_{\Dindextwo} (\pvx'' - \vx) \,
   \alpha_\pos(\ve_\Dindexfour)} (x',\px'\setminus \pvx' \cup \pvx'').
\end{align}
\end{subequations}
We read off that the first term corresponds to the jump of a single
electron from $\vx'$ to $\vx''$, while all other particles remain
where they were, the second to the annihilation of an
electron--positron pair at locations $\vx'$ and $\pvx'$, the third to
the creation of an electron--positron pair at locations $\vx'$ and
$\pvx'$, and the last to the jump of a positron from $\pvx'$ to
$\pvx''$.  The corresponding jump rates are
\begin{subequations}\label{elposrates}
\begin{align}
   \sigma_\el (x'\setminus \vx' \cup \vx'',\px'|q') &= \frac{[(2/\hbar)
   \, \Im \, \Psi^*(q) \sum_{\Dindexthree,\Dindexfour}
   \chi_\el^{\Dindexthree, \Dindexfour} (\vx',\vx'')
   \alpha_\el(\ve_\Dindexfour) \varepsilon_\el(\ve_\Dindexthree)\,
   \Psi(q')]^+}{\Psi^*(q') \, \Psi(q')} \\
   \sigma_\ann (x'\setminus \vx',\px'\setminus \pvx'|q') &=
   \frac{[(2/\hbar) \, \Im \, \Psi^*(q) \sum_{\Dindexthree,\Dindexfour}
   \chi_\ann^{\Dindexthree, \Dindexfour} (\vx',\pvx')
   \varepsilon_\pos(\ve_\Dindexfour) \varepsilon_\el(\ve_\Dindexthree)\,
   \Psi(q')]^+}{\Psi^*(q') \, \Psi(q')} \\
   \sigma_\crea (x'\cup \vx',\px' \cup \pvx'|q') &= \frac{[(2/\hbar) \,
   \Im \, \Psi^*(q) \sum_{\Dindexthree,\Dindexfour}
   \chi_\crea^{\Dindexthree, \Dindexfour} (\vx',\pvx')
   \alpha_\el(\ve_\Dindexfour) \alpha_\pos(\ve_\Dindexthree)\,
   \Psi(q')]^+}{\Psi^*(q') \, \Psi(q')} \\
   \sigma_\pos (x',\px'\setminus \pvx' \cup \pvx''|q') &=
   \frac{[(2/\hbar) \, \Im \, \Psi^*(q) \sum_{\Dindexthree,\Dindexfour}
   \chi_\pos^{\Dindexthree, \Dindexfour} (\pvx',\pvx'')
   \alpha_\pos(\ve_\Dindexfour) \varepsilon_\pos(\ve_\Dindexthree)\,
   \Psi(q')]^+}{\Psi^*(q') \, \Psi(q')},
\end{align}
\end{subequations}
where $q$ denotes the respective destination, and
\begin{subequations}
\begin{align}
   \chi_\el^{\Dindexthree, \Dindexfour}(\vx',\vx'') =\quad &
   \sum\limits_{\Dindex,\Dindextwo} \int d^3\vx \,
   {S_+}^\Dindexfour_\Dindex (\vx''-\vx) \, A^{\Dindex,\Dindextwo}
   (\vx) \, {S_+}^\Dindexthree_\Dindextwo (\vx'-\vx) \\
   \chi_\ann^{\Dindexthree, \Dindexfour} (\vx',\pvx') =\quad &
   \sum\limits_{\Dindex,\Dindextwo} \int d^3\vx \,
   {S_-}^\Dindexfour_\Dindex (\pvx'-\vx) \, A^{\Dindex,\Dindextwo}
   (\vx) \, {S_+}^\Dindexthree_\Dindextwo (\vx'-\vx) \\
   \chi_\crea^{\Dindexthree, \Dindexfour} (\vx',\pvx') =\quad
   &\sum\limits_{\Dindex,\Dindextwo} \int d^3\vx \,
   {S_+}^\Dindexfour_\Dindex (\vx'-\vx) \, A^{\Dindex,\Dindextwo} (\vx)
   \, {S_-}^\Dindexthree_\Dindextwo (\pvx'-\vx) \\
   \chi_\pos^{\Dindexthree, \Dindexfour} (\pvx',\pvx'') =
   -&\sum\limits_{\Dindex,\Dindextwo} \int d^3\vx \,
   {S_-}^\Dindexfour_\Dindextwo (\pvx''-\vx) \, A^{\Dindex,\Dindextwo}
   (\vx) \, {S_-}^\Dindexthree_\Dindex (\pvx'-\vx).
\end{align}
\end{subequations}
The process for $H_0 + H_\inter$ that we obtain through process
additivity is the motion \eqref{elposmotion} interrupted by stochastic
jumps with rates \eqref{elposrates}.

Note that the jump of a single electron has small probability to be
across a distance much larger than the width of the functions $S_\pm$,
which is of the order of the Compton wavelength of the
electron. Similarly, the distance $|\vx-\pvx|$ of a newly created
pair, or of a pair at  the moment of annihilation, has small probability
to be much larger than the width of $S_\pm$.  While the jump of a
single electron or positron leaves the number $N$ of electrons and the
number $\pN$ of positrons unchanged, pair creation and annihilation
can only either decrease or increase both $N$ and $\pN$ by $1$.  As a
consequence, the actual net charge $\pN-N$ is conserved by the
process.

\section{Second Quantization of a Markov Process}\label{sec:morefree}

\subsection{Preliminaries Concerning the Conditional Density Matrix}

In the next section, we describe the algorithm for the ``second
quantization'' of a process. But before that, we have to introduce, as
a preparation, the notion of a conditional density matrix. In
\cite{DGZ}, we have defined for Bohmian mechanics the
\emph{conditional wave function} of, say, subsystem 1 of a composite
system with configuration space $\conf = \conf_1 \times \conf_2$ by
$\Psi_\cond(q_1) = \Psi(q_1,Q_2)$.  {}From a complex wave function $\Psi
: \conf \to \CCC$, together with the actual configuration $Q_2$ of the
environment of the subsystem in the composite, we thus form a wave
function $\Psi_\cond: \conf_1 \to \CCC$; for Bohmian mechanics with
spin, in contrast, we would not, in general, obtain a suitable wave
function for subsystems in this way, because $\Psi_\cond$ as just
defined would have more spin indices than appropriate.  We can however
still define the \emph{conditional density matrix} for subsystem 1,
\begin{equation}\label{WPsi}
   W_{\cond \, s_1,s_1'}(q_1,q_1') = \frac{1}{\gamma} \sum_{s_2}
   \Psi_{s_1,s_2} (q_1,Q_2) \, \Psi^*_{s_1',s_2} (q_1', Q_2)
\end{equation}
where the $s$'s are spin indices. In order that $W$, like any
density matrix, have trace 1, the normalizing factor $\gamma$ must be
chosen as
\[
   \gamma = \int\limits_{q_1 \in \conf_1} \sum_{s_1,s_2} \Psi^*_{s_1,s_2}
   (q_1,Q_2) \, \Psi_{s_1,s_2} (q_1,Q_2) \, dq_1\,.
\]
This $W$ can play most of the roles of the conditional wave
function in spinless Bohmian mechanics.  The notion of a conditional
density matrix easily generalizes from the situation just described,
corresponding to wave functions in $L^2(\conf,\CCC^k)$ and the natural
localization PVM, to the situation of any product localization POVM on
any tensor product Hilbert space: for $\Hilbert = \Hilbert_1
\otimes \Hilbert_2$ and $\pov(dq_1 \times dq_2) = \pov_1(dq_1) \otimes
\pov_2(dq_2)$,  set
\begin{equation}\label{WPOV}
   W_\cond = \frac{\tr_2 \big( |\Psi\rangle\langle\Psi| \, \pov
   (\conf_1 \times dq_2) \big)} {\tr \big( |\Psi\rangle\langle\Psi|
   \, \pov(\conf_1 \times dq_2) \big)} \Big|_{q_2 = Q_2}\,,
\end{equation}
where $\tr_2$ is the partial trace over $\Hilbert_2$. The quotient is
to be understood as a Radon--Nikod{\'y}m derivative in $q_2$. Like
conditional wave functions, conditional density matrices cannot be
defined in orthodox quantum theory, for lack of the configuration
$Q_2$.  We stress that conditional density matrices have nothing,
absolutely nothing, to do with statistical ensembles of state vectors
in $\Hilbert_1$. Like any density matrix, they do, however, define a
probability distribution on $\conf_1$,
\begin{equation}\label{PW}
   \measure^{W_\cond}_1 (\,\cdot\,) = \tr \big(W_\cond \,
   \pov_1(\,\cdot\,) \big)\,,
\end{equation}
which coincides with the conditional distribution of $Q_1$ given $Q_2$,
\[
   \measure(Q_1 \in \,\cdot\,|Q_2) = \frac{\sp{\Psi}{\pov_1(\,\cdot\,)
   \otimes \pov_2(dq_2)| \Psi}} {\sp{\Psi}{\1 \otimes \pov_2(dq_2)|
   \Psi}} \Big|_{q_2 = Q_2}\,.
\]

The evolution of $W_\cond$ is not autonomous; it will typically depend
on (and always be determined by) $\Psi_t$ and $Q_{2,t}$. For a given
density matrix $W$ of a system that is not regarded as a subsystem,
however, one can \emph{define} (as usual) the time evolution by $W_t =
\E^{-\I H t/\hbar} \, W \, \E^{\I H t/\hbar}$, which gives rise to a
time-dependent distribution $\measure^{W_t} (\,\cdot\,) = \tr (W_t
\pov(\,\cdot\,))$. We call a Markov process that is
$\measure^{W_t}$-distributed at every time $t$ \emph{equivariant} with
respect to $W$ and $H$. Given the right initial distribution, this is
equivalent to the following condition on the generator:
\begin{equation}\label{Wequi}
   \generator \measure^W (\,\cdot\,) = \frac{2}{\hbar} \, \Im \,
   \tr(W \, \pov(\,\cdot\,) \, H)\,.
\end{equation}
This is the version of (\ref{mainequ}) for density matrices, and
defines an \emph{equivariant generator} with respect to $W$ and $H$.

Since conditional density matrices will play a crucial role in the
construction of the many-particle process, we require that, as part of
the input data of the algorithm, we are given an equivariant generator
$\generator^{(1)}_W$ for every density matrix from a dense subset of
the density matrices in $\Hilbert^{(1)*} \otimes \Hilbert^{(1)}$.
This is not much of a restriction, as all relevant examples of
equivariant generators naturally extend to density matrices: Bohmian
mechanics with spin space $\CCC^k$ can be extended \cite{Belldensity}
to
\begin{equation}\label{vW}
   v^W(q) = \hbar \, \Im \, \frac{\nabla_{q} \tr_{\CCC^k} \,
   W(q,q')}{\tr_{\CCC^k} \, W(q,q')} (q'=q)\,,
\end{equation}
Bohm--Dirac to
\begin{equation}\label{vWDirac}
   v^W(q) = \frac{\tr_{\CCC^4} (W(q,q) \valpha)}{\tr_{\CCC^4}
   (W(q,q))} \,,
\end{equation}
and minimal jump rates to
\begin{equation}\label{sigmaW}
   \sigma^W (dq|q') = \frac{[(2/\hbar)\, \Im \, \tr(W \pov(dq) H
   \pov(dq'))]^+} {\tr(W \pov(dq'))} \,.
\end{equation}
Note also that (\ref{vW}) would not make any sense if $W$ represented
a statistical ensemble \cite{Belldensity}, whereas it makes good sense
for conditional density matrices, expressing the true relation between
the Bohmian velocity for a subsystem arising from (\ref{Bohm}) and the
conditional density matrix (\ref{WPsi}) of that subsystem. Mutatis
mutandis, the same is true of (\ref{vWDirac}). Similarly, in case that
$\pov$ is a PVM, (\ref{sigmaW}) expresses the jump rates for a
decoupled subsystem arising from \eqref{tranrates} for the composite
in terms of the conditional density matrix of that subsystem.

\subsection{Algorithm}
\label{sec:Gamma}

The input data of this algorithm are the one-particle Hilbert space
$\Hilbert^{(1)}$, configuration space $\conf^{(1)}$, POVM
$\pov^{(1)}$, and a family of generators $\generator^{(1)} =
\generator^{(1)}_W$ labeled by the density matrices $W$ from a dense
subset of the density matrices in $\Hilbert^{(1)*} \otimes
\Hilbert^{(1)}$.  The output is a family of generators $\Gamma
\generator^{(1)} = \generator_0 = \generator_{0,\Psi}$ labeled by the
state vectors $\Psi$ in (a dense subspace of) Fock space. If
$\generator^{(1)}_W$ is equivariant with respect to $W$ and $H^{(1)}$,
then $\generator_{0,\Psi}$ is equivariant with respect to $\Psi$ and
$H_0$.

The algorithm is based on two procedures for suitably combining
generators for direct sums or tensor products of Hilbert spaces.

\subsubsection{Direct Sums}\label{sec:directsum}

   Given a finite or countable sequence of Hilbert
   spaces $\Hilbert^{(n)}$ with POVMs $\pov^{(n)}$ on configuration
   spaces $\conf^{(n)}$, and for each $n$ a family of generators
   $\generator^{(n)}$ labeled by the vectors in $\Hilbert^{(n)}$, there
   is a canonically constructed family of generators $\generator^\oplus
   = \generator^\oplus_\Psi$, labeled by the vectors in the direct sum
   $\bigoplus_n \Hilbert^{(n)}$. The space $\conf$ in which the
   corresponding process takes place is the disjoint union of the
   $\conf^{(n)}$. If every $\generator^{(n)}_{\Psi_n}$ is equivariant
   with respect to $\Psi_n \in \Hilbert^{(n)}$ and $H^{(n)}$, then
   $\generator^\oplus_\Psi$ is equivariant with respect to $\Psi \in
   \bigoplus_n \Hilbert^{(n)}$ and $\bigoplus_n H^{(n)}$.

   Here are the details. The POVM $\pov = \bigoplus_n \pov^{(n)}$ on
   $\conf$ that naturally arises from the data is given by $\pov(B) =
   \bigoplus_n \pov^{(n)} (B \cap \conf^{(n)})$ for $B \subseteq
   \conf$.  Let $P_n$ denote the projection $\Hilbert \to
   \Hilbert^{(n)}$.  The generator $\generator^\oplus$ is given by
   \begin{equation}
     \big( \generator_\Psi^\oplus \, \rho \big) \big|_{\conf^{(n)}} =
     \generator_{P_n\Psi/\|P_n\Psi\|}^{(n)} \big(
     \rho \big|_{\conf^{(n)}} \big)\,.
   \end{equation}
   It generates a (Markov) process $Q_t^\oplus$ such that when
   $Q_0^\oplus \in \conf^{(n)}$, it is generated by the state vector
   $P_n \Psi/ \|P_n \Psi \|$, i.e., it is a Markov process $Q_t^{(n)}$
   in $\conf^{(n)}$ generated by $\generator^{(n)}_{P_n \Psi/
   \|P_n\Psi\|}$.  The equivariance statement follows directly, since
   $\|P_n \Psi_t\|^2 = \measure_t (\conf^{(n)})$  is invariant under the
   evolution generated by $H_0 = \bigoplus_n H^{(n)}$.

\subsubsection{Tensor Products}\label{sec:tensorproduct}

   Given a finite sequence of Hilbert spaces
   $\Hilbert^{[1]}, \ldots, \Hilbert^{[n]}$ with POVMs $\pov^{[i]}$ on
   configuration spaces $\conf^{[i]}$, and for each $i$ a family of
   generators $\generator^{[i]} = \generator^{[i]}_{W_i}$  labeled
   by the density matrices on $\Hilbert^{[i]}$, there is a canonically
   constructed family of generators $\generator^\otimes =
   \generator^\otimes_W$, labeled by the density matrices on the tensor
   product $\Hilbert^{[1]} \otimes \cdots \otimes \Hilbert^{[n]}$. The
   corresponding process takes place in the Cartesian product $\conf =
   \conf^{[1]} \times \cdots \times \conf^{[n]}$. If every
   $\generator^{[i]}_{W_i}$ is equivariant with respect to the density
   matrix $W_i$ on $\Hilbert^{[i]}$ and the Hamiltonian $H^{[i]}$, then
   $\generator^\oplus_W$ is equivariant with respect to $W$ on
   $\Hilbert^{[1]} \otimes \cdots \otimes \Hilbert^{[n]}$ and $H =
   \sum\limits_i \1 \otimes \cdots \otimes H^{[i]} \otimes \cdots
   \otimes \1 = \sum\limits_i H_i$.

\newcommand{\hqi}{\widehat{q}_i}

   Here are the details. The POVM that naturally arises from the data
   is\footnote{The existence of the tensor product POVM is a
   consequence of Corollary~7 in Section~4.4 of \cite{crea2A}.}
   \begin{equation}\label{productpovm}
     \pov(d\vq_1 \times \cdots \times d\vq_n) = \pov^{[1]}(d\vq_1)
     \otimes \cdots \otimes \pov^{[n]}(d\vq_n).
   \end{equation}
   For any $q \in \conf$, let $\vq_i$ denote its $i$-th component and
   let $\hqi = (\vq_1, \ldots, \vq_{i-1}, \vq_{i+1}, \ldots,
   \vq_n)$. For every $i$ and $\hqi$, define
   \[
     W_i (\hqi) = \frac{\tr_{\neq i} \big( W \pov(d\vq_1 \times \cdots
     \times \conf^{[i]} \times \cdots \times d\vq_n) \big)} {\tr \big(W
     \pov(d\vq_1 \times \cdots \times \conf^{[i]} \times \cdots \times
     d\vq_n) \big)}\,,
   \]
   where $\tr_{\neq i}$ is the partial trace over all factors except
   $\Hilbert^{[i]}$. This $W_i$ is the conditional density matrix,
   regarded as a function of the configuration $\hqi$ of the other
   particles.  Now consider the process on $\conf$ according to which
   the $i$-th particle moves as prescribed by $\generator^{[i]}_{W_i}$
   while the other particles remain fixed.  The generator of this
   process is
   \begin{equation}\label{Lidef}
     \generator_i \, \rho := \Big[ \generator^{[i]}_{W_i(\hqi)} \,
     \rho( \,\cdot\,| \hqi) \Big] \, \rho_{\neq i}(d\hqi)
   \end{equation}
   where $\rho_{\neq i}$ is the marginal distribution of
   $\widehat{Q}_i$ (i.e., $\rho$ integrated over $\vq_i$) and
   $\rho(\,\cdot\,|\hqi)$ is the conditional distribution of $\vQ_i$
   given $\widehat{Q}_i = \hqi$; the square bracket is a function of
   $\hqi$ and a measure in $d\vq_i$.  Now define $\generator^\otimes_W
   \rho = \sum\limits_i \generator_i \rho$.

   To see that $\generator^\otimes$ is equivariant when the
   $\generator^{[i]}$ are, we have to check (\ref{Wequi}).  Note first
   that $\measure^W(d\vq_i|\hqi) = \tr \big( W_i(\hqi) \,
   \pov^{[i]}(d\vq_i) \big)$.  Due to the equivariance of
   $\generator^{[i]}$, for $\rho = \measure^W$ the square bracket in
   (\ref{Lidef}) equals $(2/\hbar) \, \Im \, \tr \big( W_i (\hqi) \,
   \pov^{[i]}(d\vq_i) \, H^{[i]} \big)$, from which we obtain
   (\ref{Wequi}) for $\generator_i$ and $H_i$ and hence for
   $\generator^\otimes$ and $H$.

The definition of $\generator^\otimes$ reproduces the many-particles
Bohm law (\ref{Bohm}) with or without spin from the one-particle
version (or, for distinguishable particles, from several different
one-particle versions having different masses and spins). Similarly,
it reproduces the many-particles Bohm--Dirac law (\ref{BohmDirac})
from the one-particle version.

\subsubsection{Second Quantization of the POVM}\label{sec:GammaPOVM}

Let $\conf^{(n)}$ denote the space of all subsets-with-multiplicities
of $\conf^{(1)}$ having $n$ elements (counting in the multiplicities).
$\pov^{(1)}$ naturally defines a POVM $\pov^{(1)\otimes n}$ on
$(\conf^{(1)})^n$ acting on $\Hilbert^{(1)\otimes n}$ by
$\pov^{(1)\otimes n}(d\vq_1 \times \cdots \times d\vq_n) =
\pov^{(1)}(d\vq_1) \otimes \cdots \otimes \pov^{(1)}(d\vq_n)$, and a
POVM $\pov^{(n)}$ on $\conf^{(n)}$ acting on $\Fock^{(n)} = P_\pm
\Hilbert^{(1)\otimes n}$ (the $n$-particle sector of Fock space, with
$P_\pm$ the projection to the subspace of (anti\nobreakdash-)symmetric
elements of $\Hilbert^{(1)\otimes n}$, depending on whether we deal
with fermions or bosons) by
\[
   \pov^{(n)}(B) = \pov^{(1)\otimes n} \big\{(\vq_1, \ldots, \vq_n)
   \in (\conf^{(1)})^n : \{\vq_1, \ldots, \vq_n\} \in B \big\}
\]
for $B \subseteq \conf^{(n)}$, where $\{\vq_1, \ldots, \vq_n\}$ should
be understood as a set-with-multiplicities.\footnote{This agrees with
the definition given in Section \ref{sec:crea1} for the case of a PVM
and the coincidence configurations removed from configuration space.}
Since $\pov^{(n)}(B)$ is invariant under permutations, it maps
symmetric to symmetric and anti-symmetric to anti-symmetric elements
of $\Hilbert^{(1)\otimes n}$ and thus acts on $\Fock^{(n)}$ for
bosonic or fermionic Fock space.\footnote{In case that $\pov^{(1)}$ is
nonatomic, $\pov^{(n)}$ can equivalently be defined in the following
way: For the set $\Delta$ of coincidence configurations we set
$\pov^{(n)}(\Delta) =0$, and for volumes $d\vq_1, \ldots, d\vq_n$ in
$\conf^{(1)}$ that are pairwise disjoint, we have a corresponding
volume $dq$ in $\conf^{(n)}$, which can be obtained from $d\vq_1
\times \cdots \times d\vq_n \subseteq (\conf^{(1)})^n$ by forgetting
the ordering, and we  set $\pov^{(n)}(dq) = n! \, P_\pm \,
\pov^{(1)}(d\vq_1) \otimes \cdots \otimes \pov^{(1)}(d\vq_n) \,
P_\pm$.} The corresponding POVM on $\conf$ is then $\pov = \Gamma
\pov^{(1)} = \bigoplus_n \pov^{(n)}$; more precisely, for $B \subseteq
\conf$,
\[
   \pov(B) = \bigoplus_{n=0}^\infty \pov^{(n)} (B\cap \conf^{(n)})\,.
\]

\subsubsection{Construction of the Free Process}

Equipped with the two procedures for direct sums and tensor products,
we complete the construction of the free process.

The ``tensor product'' procedure above provides a process on
$(\conf^{(1)})^n$ from $n$ identical copies of $\generator^{(1)}$.
For a state vector $\Psi^{(n)} \in \Fock^{(n)} = P_\pm
\Hilbert^{(1)\otimes n}$ from either the symmetric or the
anti-symmetric elements of the $n$-fold tensor product space, let $W$
be the projection to $\Psi^{(n)}$; the generator
$\generator^\otimes_W$ is permutation invariant because the
tensor-product construction of $\generator^\otimes _W$ is permutation
covariant and a permutation can at most change the state vector by a
minus sign, which does not affect the density matrix. Consequently,
the ordering of the configuration is irrelevant and may be ignored. We
thus obtain a process on $\conf^{(n)}$ whose generator we call
$\generator^{(n)}$. We now apply the ``direct sum'' procedure to
obtain a process on $\conf$.

\section{Towards a Notion of Minimal Process}

In this section, we investigate the common traits of the Markov
processes relevant to Bell-type QFT, which can be summarized in the
notion of a \emph{minimal process} associated with $\Psi,H$, and
$\pov$. We begin with a closer study of the minimal free generator
\eqref{LH}, and then explain  why we call the minimal jump rates
``minimal.'' Finally, in Section \ref{sec:mini}, we give an outlook on
the notion of minimal process.

\subsection{Free Process From Differential Operators}
\label{sec:freeflow}

In this section, we discuss some of the details, concerning the two
equivalent formulas \eqref{LH} and \eqref{genH} for the backward and
forward version of the minimal free generator in terms of $H, \pov$,
and $\Psi$, that we omitted in Section \ref{sec:free2}.  To begin
with, $L$ as defined by \eqref{LH} satisfies some necessary conditions
for being a backward generator: $Lf(q)$ is real, and $L\1 =0$ where
$\1$ is the constant 1 function (this corresponds to $\generator \rho
(\conf) =0$, or conservation of total probability). In case $L$ is
indeed a backward generator, the corresponding process is equivariant
because
\[
   \generator \measure (dq) \stackrel{\eqref{genH}}{=} \Re \, \sp{\Psi}
   {\hat{\1}\,\frac{\I}{\hbar} [H,\pov(dq)] |\Psi} = \frac{2}{\hbar} \,
   \Im \, \sp{\Psi} {\pov(dq) H|\Psi} \stackrel{\eqref{dPdt}}{=}
   \dot{\measure}(dq)\,.
\]

  One way to arrive at formula \eqref{LH} has been described in Section
\ref{sec:free2}.  A different way, leading to \eqref{genH}, is to
start from the ansatz $\generator \rho = A\frac{d\rho}{d\measure}$
where $A$ denotes a (signed-measure-valued) linear operator acting on
functions.  Equivariance means $A\1 (dq) = \sp{\Psi}{\frac{\I}{\hbar}
[H, \pov(dq)] |\Psi}$.  This suggests $Af(dq) = \sp{\Psi}{\hat{f}\,
\frac{\I}{\hbar} [H, \pov(dq)] |\Psi}$, or $Af(dq) =
\sp{\Psi}{\frac{\I}{\hbar} [H, \pov(dq)]\, \hat{f} |\Psi}$, or a
convex combination thereof.  Since $Af(dq)$ must be real, we are
forced to choose the combination with coefficients $\frac{1}{2}$ and
$\frac{1}{2}$, or equivalently $Af(dq) = \Re\, \sp{\Psi}{\hat{f}\,
\frac{\I}{\hbar} [H, \pov(dq)] |\Psi}$, which is \eqref{genH}.

That $\generator$ generates a deterministic process (when it is a
generator at all) is suggested by the following consideration---at
least when $H$ and $\pov$ are time-reversal invariant: replacing
$\Psi$ in \eqref{genH} by $T\Psi$ where $T$ is the anti-linear time
reversal operator (see Section \ref{sec:symm}) changes the sign of
$\generator$. The only generators $\generator$ such that $-\generator$
is also a generator are, presumably, those corresponding to
deterministic motion.

This gives us an opportunity to check for which $H$ \eqref{LH} does
define a process: for a deterministic process we must have $L = v\cdot
\nabla$ where $v$ is the velocity vector field.  It is known that
vector fields, understood as first-order differential operators, are
those linear operators $L$ on the space of smooth functions that
satisfy the Leibniz rule $L(fg) = fLg + gLf$.  \eqref{LH} is certainly
linear in $f$, so we have to check the Leibniz rule to see whether $L$
is indeed of the form $v\cdot \nabla$ and thus the backward generator
of a process.

We can see no reason why $L$ should satisfy a Leibniz rule unless
$\pov$ is a PVM, which implies that
\begin{equation}\label{fPOV}
   \hat{f} \, \pov(dq) = f(q) \, \pov(dq)\,,
\end{equation}
and $H$ is such that for all (nice) functions $f$ and $g$,
\begin{equation}\label{Hdiff}
   \big[ [ H,\hat{f}] , \hat{g} \big] = \hat{h}
\end{equation}
for some function $h$, which holds if $H$ is a differential operator
of order $\leq 2$. (If $H=-\Laplace$, then $h=- 2 \nabla f \cdot
\nabla g$; if $H=-\I \, \valpha \cdot \nabla$ for whatever vector of
matrices $\valpha$, or if $H$ is a multiplication operator, then
$h=0$.)  To check that the Leibniz rule is obeyed in this case, note
that we then have  that $[H, \widehat{fg}] = [H, \hat{f} \hat{g}] =
[H,\hat{f}] \hat{g} + \hat{f} [H,\hat{g}] = \hat{f} [H, \hat{g}] +
\hat{g} [H, \hat{f}] + \big[ [H, \hat{f}], \hat{g} \big]$.  Using this
in \eqref{LH}, we find that, due to \eqref{fPOV}, the first two terms
give the Leibniz rule, whereas the last term, due to \eqref{Hdiff},
does not contribute to the real part in \eqref{LH}.

When $\Hilbert$ is an $L^2$ space over $\conf$ and $\pov$ the natural
PVM, i.e., when $\Psi$ is a function, \eqref{LH} can be written in the
form
\begin{equation}\label{vH}
   L f(q)= \frac{1}{\hbar}\, \Im \, \frac{\Psi^*(q) \,
   ([\hat{f},H]\Psi)(q)} {\Psi^*(q) \, \Psi(q)}
\end{equation}
where $\hat{f}$ is the multiplication operator corresponding to $f$.
{}From this, one easily reads off the Bohm velocity \eqref{Bohm} for the
$N$-particle Schr\"odinger operator \eqref{Hamil} with or without
spin.  Similarly, we get the Bohm--Dirac theory when $H$ is the Dirac
operator in $\Hilbert = \Anti L^2(\RRR^3,\CCC^4)^{\otimes N}$, $\conf$
the manifold of subsets of $\RRR^3$ with $N$ elements, and $\pov$ the
obvious PVM.  \eqref{vH} also leads to the Bohm--Dirac motion if
$\Hilbert = L^2(\RRR^3,\CCC^4)^{\otimes N}$, $\conf = \RRR^{3N}$, and
$\pov$ is the natural PVM, but not if $\Hilbert$ is the positive
energy subspace because then the appropriate POVM $\pov$ is no longer
a PVM.

To see that the ``second quantization'' algorithm maps minimal free
generators to minimal free generators, or, in other words, preserves
the relation \eqref{genH} between Hamiltonian and generator, observe
first that \eqref{genH} naturally extends to density matrices, and the
extension, if a generator, is equivariant. Next check that the
``direct sum'' and ``tensor product'' procedures of Section
\ref{sec:Gamma} are compatible with \eqref{genH} when $\pov$ is a PVM.
Finally, observe that the (anti\nobreakdash-)symmetrization operator
commutes with the $n$-particle Hamiltonian, with $\pov(B)$ for every
permutation invariant set $B \subseteq (\conf^{(1)})^n$, and with
$\hat{f}$ for every permutation invariant function $f:(\conf^{(1)})^n
\to \RRR$.

\subsection{Minimality}
\label{sec:mini4}

In this section we explain in what sense the minimal jump rates
\eqref{tranrates}---or \eqref{mini1}---are minimal.  In so doing, we
will also explain the significance of the quantity $\current$ defined
in \eqref{Jdef}, and clarify the meaning of the steps taken in
Sections \ref{sec:mini1} and \ref{sec:mini2} to arrive at the jump
rate formulas.

Given a Markov process $Q_t$ on $\conf$, we define the \emph{net
probability current} $j_t$ at time $t$ between sets $B$ and $B'$ by
\begin{eqnarray}\label{jdefcont}
     j_t(B,B') = \lim_{\Delta t \searrow 0} \frac{1}{\Delta t}
\hspace{-3ex}
   &&
     \Big[ \prob\big\{Q_{t}\in B',Q_{t+\Delta t}\in B  \big\} -
   \\\nonumber
   &&
   - \prob \big\{ Q_{t}\in B, Q_{t+\Delta t} \in B' \big\} \Big]\,.
\end{eqnarray}
This is the amount of probability that flows, per unit time, from $B'$
to $B$ minus the amount from $B$ to $B'$.  For a pure jump process, we
have that
\begin{equation}\label{jrate}
   j_t(B,B') = \int\limits_{q'\in B'} \sigma_t(B|q')\, \rho_t(dq') -
   \int\limits_{q\in B} \sigma_t(B'|q)\, \rho_t(dq)\,,
\end{equation}
so that
\begin{equation}
j_t(B,B') = j_{\sigma,\rho}(B \times B')
\end{equation}
where $j_{\sigma,\rho}$ is the signed measure, on $\conf \times
\conf$, given by the integrand of \eqref{continuity3},
\begin{equation}\label{jsigma}
   j_{\sigma,\rho} (dq \times dq') = \sigma(dq|q') \, \rho(dq') -
   \sigma(dq'|q) \, \rho(dq)\,.
\end{equation}
For minimal jump rates $\sigma$, defined by \eqref{tranrates} or
\eqref{mini1} (and with the probabilities $\rho$ given by \eqref{mis},
$\rho = \measure$), this agrees with \eqref{Jdef}, as was noted
earlier,
\begin{equation}\label{jJ}
   j_{\sigma,\rho} = \current_{\Psi,H,\pov} \,,
\end{equation}
where we have made explicit the fact that $\current$ is defined in
terms of the quantum entities $\Psi, H$, and $\pov$. Note that both
$\current$ and the net current $j$ are anti-symmetric, $\current^\tr =
-\current$ and $j^\tr = -j$, the latter by construction and the former
because $H$ is Hermitian. (Here $\tr$ indicates the action on measures
of the transposition $(q,q') \mapsto (q',q)$ on $\conf \times \conf$.)
The property \eqref{jJ} is stronger than the equivariance of the rates
$\sigma$, $\generator_\sigma \measure_t = d\measure_t / dt$: Since, by
\eqref{continuity3},
\begin{equation}
   (\generator_\sigma \rho) (dq) = j_{\sigma,\rho} (dq \times \conf),
\end{equation}
and, by \eqref{Jdef},
\begin{equation}
   \frac{d\measure}{dt}(dq) = \current(dq \times \conf),
\end{equation}
the equivariance of the jump rates $\sigma$ amounts to the condition
that the marginals of both sides of \eqref{jJ} agree,
\begin{equation}
   j_{\sigma,\rho} (dq \times \conf) = \current (dq \times \conf)\,.
\end{equation}
In other words, what is special about processes with rates satisfying
\eqref{jJ} is that not only the single-time \emph{distribution} but
also the \emph{current} is given by a standard quantum theoretical
expression in terms of $H, \Psi$, and $\pov$. That is why we call
\eqref{jJ} the \emph{standard-current property}---defining
\emph{standard-current rates} and \emph{standard-current processes}.

Though the standard-current property is stronger than equivariance, it
alone does not determine the jump rates, as already remarked in
\cite{BD,Roy}. This can perhaps be best appreciated as follows: Note
that \eqref{jsigma} expresses $j_{\sigma,\rho}$ as twice the
anti-symmetric part of the (nonnegative) measure
\begin{equation}
  C(dq \times dq') = \sigma(dq|q') \, \rho(dq')
\end{equation}
on $\conf \times \conf$ whose right marginal $C(\conf \times dq')$ is
absolutely continuous with respect to $\rho$. Conversely, from any
such measure $C$ the jump rates $\sigma$ can be recovered by forming
the Radon--Nikod\'ym derivative
\begin{equation}
   \sigma(dq|q') = \frac{C(dq \times dq')}{\rho(dq')}\,.
\end{equation}
Thus, given $\rho$, specifying $\sigma$ is equivalent to specifying
such a measure $C$.

In terms of $C$, the standard-current property becomes (with $\rho =
\measure$)
\begin{equation}\label{CJ}
   2 \, \mathrm{Anti} \, C = \current.
\end{equation}
Since (recalling that $\current = \current^+ - \current^-$ is
anti-symmetric)
\begin{equation}
   \current = 2 \, \mathrm{Anti} \, \current^+,
\end{equation}
an obvious solution to \eqref{CJ} is
\[
   C = \current^+,
\]
corresponding to the minimal jump rates. However, \eqref{jJ} fixes
only the anti-symmetric part of $C$. The general solution to
\eqref{CJ} is of the form
\begin{equation}
   C = \current^+ + S
\end{equation}
where $S(dq \times dq')$ is symmetric, since any two solutions to
\eqref{CJ} have the same anti-symmetric part, and $S \geq 0$, since $S
= C \wedge C^\tr$, because $\current^+ \wedge (\current^+)^\tr =0$.

In particular, for any standard-current rates, we have that
\begin{equation}\label{minimality}
   C \geq \current^+, \quad \text{or} \quad \sigma(dq|q') \geq
   \frac{\current^+(dq \times dq')}{\measure(dq')}.
\end{equation}
Thus, among all jump rates consistent with the standard-current
property, one choice, distinguished by equality in \eqref{minimality},
has the least frequent jumps, or the smallest amount of stochasticity:
the minimal rates \eqref{tranrates}.

\subsection{Minimal Processes}
\label{sec:mini}

We have considered in this paper minimal jump processes, i.e., jump
processes with rates \eqref{tranrates}, associated with integral
operators $H$. There is a more general notion of minimal process, such
that there is a minimal process associated with every Hamiltonian from
a much wider class than that of integral operators; a class presumably
containing all Hamiltonians relevant to QFT. This will be discussed in
detail in a forthcoming work \cite{crea3}.

Bohmian mechanics is, in this sense, the minimal process associated with
the Schr\"odin\-ger Hamiltonian \eqref{Hamil}. The minimal process
associated
with an integral operator is the jump process with minimal rates. When 
the
minimal free generator \eqref{LH} exists, i.e., when \eqref{LH} is a
generator, it generates the minimal process associated with $H$. The
minimal process associated with the Hamiltonian of a QFT is the one we
have
obtained in this paper by means of process additivity. The concept of
minimal process directly provides, perhaps always,  the process relevant
to a Bell-type QFT.

To  begin to convey the notion of the minimal process, we generalize
the standard-current property (cf.\ Section \ref{sec:mini4}) from pure
jump processes to general Markov processes: the net probability
current $j$ of a Markov process defines a bilinear form
\begin{equation}
   j_t(f,g) = \lim_{\Delta t \searrow 0} \, \frac{1}{\Delta t} \, \EEE
   \big( f(Q_{t+\Delta t}) g(Q_t) - f(Q_t) g(Q_{t + \Delta t}) \big)
   = (g,L_t f) - (f, L_t g)
\end{equation}
where $L_t$ is its backward generator, and $( \;, \, )$ on the right
hand side means the scalar product of $L^2(\conf, \rho_t)$.  Then the
Markov process satisfies the \emph{standard-current property} if
$\rho_t = \measure_t$ and (for $f$ and $g$ real) $j_t(f,g)$ is equal
to
\begin{equation}
   \current_t(f,g) = \frac{2}{\hbar} \, \Im \, \sp{\Psi_t} {\hat{f} H
   \hat{g} |\Psi_t}\,,
\end{equation}
or, in other words, if twice the anti-symmetric part of its backward
generator $L_t$ agrees with the operator corresponding to $\current_t$
as given by $(\current_t f,g) = \current_t(f,g)$, $2 \, \mathrm{Anti}
\, L_t = \current_t$.  The minimal process is then the
standard-current process that has, in a suitable sense, the smallest
amount of randomness.

Let us consider some examples. The diffusion process with generator
$\generator$ given below (and for $\rho = \measure$) has the
standard-current property (in fact, because its ``current velocity''
\cite{stochmech} is $v$)   for the Schr\"odinger Hamiltonian
\eqref{Hamil} but is not minimal:
\begin{equation}\label{diffusion}
   \generator \rho= \frac{\lambda}{2} \Laplace \rho -\div (\rho
   \tilde{v}),\mbox{ with } \tilde{v}:= v + \frac{\lambda}{2}
   \nabla(\log|\Psi|^2)
\end{equation}
where $\lambda$ is any positive constant (the diffusion constant) and
$v$ is the Bohmian velocity (\ref{Bohm}); this process was already
considered in \cite{Jaekel,Davidson}.  Note that Nelson's stochastic
mechanics \cite{stochmech} corresponds to $\lambda=\hbar$. It is
obvious without any mathematical analysis that the smallest amount of
stochasticity corresponds to absence of diffusion, $\lambda =0$, which
yields Bohmian mechanics.  Processes like the diffusion
(\ref{diffusion}) for $\lambda > 0$ seem less natural for the
fundamental evolution law of a physical theory since they involve
greater mathematical complexity than is needed for a straightforward
association of a process with $H$ and $\Psi$.  Examples of processes
that do not have the standard-current property, for the Schr\"odinger
Hamiltonian \eqref{Hamil}, are provided by the alternative velocity
formulas considered by Deotto and Ghirardi \cite{Deotto}; one can say
that their current is not the one suggested by $H$ and $\Psi$.

We return to the general discussion of the minimal process. As we have
already indicated, when, for a standard-current process, we view
$\current$ as well as its backward generator $L$ as operators on
$L^2(\conf, \measure)$, then $\frac12 \current$ is the anti-symmetric
(skew-adjoint) part of $L$; thus, only the symmetric (self-adjoint)
part of $L$ remains at our disposal.  Since one of the properties of a
backward generator   is $L\1 =0$, the first possibility $\tilde{L}$ for
$L$ that may satisfy the formal criteria for being a backward
generator is $\tilde{L} f = \frac12 \current f - (\frac12 \current
\1)f$.  When $\pov$ is a PVM, this is also the operator we obtain by
applying, to an arbitrary quantum Hamiltonian $H$, the formula
\eqref{LH} for what we called the minimal free generator, which we
repeat here for convenience:
\begin{equation}\label{Ltilde}
   \tilde{L} f(q) = \Re \, \frac{\sp{\Psi} {\pov(dq) \frac{\I}{\hbar}
   [H,\hat{f}] |\Psi}} {\sp{\Psi} {\pov(dq)|\Psi}}\,.
\end{equation}
Whereas this formula merely provided an alternative definition of the
free process in Section \ref{sec:free2}, it now plays a different
role: a step towards obtaining the minimal process from the
Hamiltonian $H$.  As we have pointed out in Section \ref{sec:free2},
$\tilde{L}$ is also an obvious naive guess for the backward generator
$L$, quite independent of equivariance or the current $\current$,
since $\frac{\I} {\hbar} [H,\hat{f}]$ is the time derivative of
$\hat{f}$. Moreover, it manifestly satisfies $\tilde{L} \1 =0$.  For
the backward generator $L$ of a standard-current process we must have,
when $\pov$ is a PVM, that $L = \tilde{L} + S$ where $S$ is a
symmetric operator and $S\1 =0$.  For the minimal process, we have to
choose $S$ as small as possible---while keeping $S$ symmetric and $L$
a backward generator.

Suppose $\pov$ is a PVM. Observe then that if $H$ is a differential
operator (as $H_0$ often is) of the kind considered in Section
\ref{sec:free2}, $\tilde{L}$ is itself a backward generator, so that
$S=0$ is a possible, and in fact the smallest, choice.  If $H$ is an
integral operator, what keeps $\tilde{L}$, an integral operator as
well, from being a backward generator is that the off-diagonal part of
its $\measure$-kernel $(q,\tilde{L} q') = \measure(q) \tilde{L}(q,q')
= \frac{1}{\hbar} \, \Im \, \sp{\Psi}{q} \sp{q}{H|q'} \sp{q'}{\Psi}$
may assume negative values whereas the off-diagonal part of the
$\measure$-kernel of $L$, $(q,Lq') = \measure(q) \sigma(q|q')$, cannot
be negative.  The smallest possible choice of $S$ has as off-diagonal
elements what is needed to compensate the negative values, and this
leads to the minimal jump process, as described in Section
\ref{sec:mini4}.  The diagonal part contains only what is needed to
ensure that $S\1 =0$.  For $H$ of the form $H_0 + H_\inter$, the role
of $S$ is again to compensate negative values off the diagonal, and
the minimal process has velocities determined by $H_0$ via \eqref{LH}
and jump rates determined by $H_\inter$ via \eqref{tranrates}.

In any case, the backward generator of the minimal process is the one
closest, in a suitable sense, to \eqref{Ltilde}.  This formula may
thus be regarded as containing the essential structure of $L$, for the
deterministic as well as for the jump part of the process.

Another approach towards a general notion of minimal process may be to
approximate $H$ by Hilbert--Schmidt operators $H_n$, with which are
associated, according to the results of Sections~4.2.1 and 4.2.4 of
\cite{crea2A}, minimal jump processes $Q_n$, and take the limit $n \to
\infty$ of the processes $Q_n$. This leads to a number of mathematical
questions, such as under what conditions on $H, \Psi, \pov$, and $H_n$
does a limiting process exist, and is it independent of the choice of
the approximating sequence $H_n$.

\section{Remarks}\label{sec:remarks}

\subsection{Symmetries}\label{sec:symm}

Process additivity preserves symmetries, in the sense that the process
generated by $\sum \generator^{(i)}$ shares the symmetries respected
by all of the building blocks $\generator^{(i)}$.  This section
elaborates on this statement, and the following ones: The minimal jump
rates \eqref{tranrates} and the minimal free generator \eqref{LH}
share the symmetries of the Hamiltonians with which they are
associated. The ``second quantization'' algorithm preserves the
symmetries respected by the one-particle process.

Here are some desirable symmetries that may serve as examples: space
translations, rotations and inversion, time translations and
reversal, Galilean or Lorentz boosts, global change of phase $\Psi \to
\E^{\I\theta} \Psi$, relabeling of particles,\footnote{This may mean
two things: changing the artificial labels given to identical
particles, or exchanging two species of particles.}  and gauge
transformations.

We focus first on symmetries that do not involve time in any way,
such as rotation symmetry. In this case, a symmetry group $G$ acts on
$\conf$, so that to every $g \in G$ there corresponds a mapping
$\varphi^g:
\conf \to \conf$.  In addition, $G$ acts on $\Hilbert$ through a
projective unitary (or anti-unitary) representation, so that to every $g
\in G$ there corresponds a unitary (or anti-unitary) operator $U_g$.
Then the theory is $G$-invariant if both the wave function dynamics
and the process on $\conf$ are, i.e., if $H$ is $G$-invariant,
\begin{equation}\label{HGinv}
   U_g^{-1} H U_g = H\,,
\end{equation}
and
\begin{equation}\label{QGinv}
   \varphi^g(Q_t^\Psi) = Q_t^{U_g \Psi}
\end{equation}
in distribution on path space.  A necessary condition for
(\ref{QGinv}) is that the ``configuration observable'' transforms like
the configuration, in the sense that
\begin{equation}\label{povGinv}
   U_g^{-1} \pov(\,\cdot\,) U_g = \varphi^g_* \pov(\,\cdot\,)\,,
\end{equation}
where $\varphi_*$ denotes the action of $\varphi$ on measures.
Without (\ref{povGinv}), (\ref{QGinv}) would already fail at time
$t=0$, no matter what the generator is.  Given (\ref{povGinv}),
(\ref{QGinv}) is equivalent to the $G$-invariance of the generator:
\begin{equation}\label{LGinv}
   \varphi^g_* \generator^\Psi \varphi^{g^{-1}}_* =
   \generator^{U_g \Psi} \,.
\end{equation}
Since $\varphi^g_*$ is a linear operator, it follows immediately that
the sum of $G$-invariant generators is again $G$-invariant.  The
minimal jump process, when it exists, is $G$-invariant, as follows
from the fact that $\varphi^g_*\sigma^\Psi(dq|\varphi^g(q')) =
\sigma^{U_g \Psi} (dq|q')$, which can be seen by inspecting the jump
rate formula (\ref{tranrates}). The minimal free generator
\eqref{genH} satisfies \eqref{LGinv} by virtue of \eqref{HGinv} and
\eqref{povGinv}. ``Second quantization'' provides $G$-actions on
$\Gamma \conf^{(1)}$ and $\Fock = \Gamma \Hilbert^{(1)}$ from given
actions on $\conf^{(1)}$ and $\Hilbert^{(1)}$; \eqref{HGinv},
\eqref{povGinv} and \eqref{LGinv} are inherited from their 1-particle
versions.

Time-translation invariance is particularly simple. Consider
generators $\generator^{(i)}_\Psi$ which do not depend on time except
through their dependence on $\Psi$.  Then the same is true of $\sum
\generator^{(i)}$. The same can be said of the ``second quantized''
generator, and, provided $H$ is time-independent, of the minimal jump
rates (\ref{tranrates}) and the minimal free generator \eqref{genH}.

Next we consider time reversal.  It is represented on $\Hilbert$ by an
anti-unitary operator $T$, i.e., an anti-linear operator such that
$\sp{T\Phi}{T\Psi}$ is the conjugate of $\sp{\Phi}{\Psi}$. We assume
that the Hamiltonian is reversible, $THT^{-1} = H$. Then the
reversibility of the theory means that
\begin{equation}\label{QTinv}
   Q^\Psi_{-t} = Q_t^{T\Psi}
\end{equation}
in distribution on path space, where the superscript should be
understood as indicating the state vector at $t=0$.  The necessary
condition analogous to (\ref{povGinv}) reads
\begin{equation}\label{povTinv}
   T^{-1} \pov(\,\cdot\,) T = \pov(\,\cdot\,) \,,
\end{equation}
and given that, (\ref{QTinv}) is equivalent to the $T$-invariance of
the generator:
\begin{equation}\label{LTinv}
   \ov{\generator}_\Psi = \generator_{T\Psi}\,, \mbox{ or }
   \ov{L}_\Psi = L_{T\Psi}\,,
\end{equation}
where $\ov\generator$ and $\ov{L}$ denote the forward and backward
generator of the time-reversed process.  $\ov{L}$ can be computed from
$L$,
for an equivariant Markov process, according to\footnote{To make this
formula plausible, it may be helpful to note that the second term on
the right hand side is just the correction needed to ensure that
$L^\dag\1 =0$, a necessary condition for being a backward generator.
If $\measure$ were stationary, the second term on the right hand side
would vanish.

   Here is a derivation of \eqref{LT}: Let $(f,g) = \int_{q\in \conf}
   f(q) \, g(q) \, \measure(dq)$ be the scalar product in
   $L^2(\conf,\measure)$. It follows from the definition
   \eqref{backgenerator} of $L$ that
   \[
     (g,Lf) = \lim_{t\searrow 0} \frac{1}{t} \, \EEE \big( g(Q_0)
     f(Q_t) - g(Q_0) f(Q_0) \big)\,.
   \]
   Correspondingly, $\ov{L}$ is characterized (for $f$ and $g$ real)  by
   \begin{eqnarray*}
     (g,\ov{L}f)
     &=& \lim_{t\searrow 0} \frac{1}{t} \, \EEE \big( g(Q_0)
     f(Q_{-t}) - g(Q_0) f(Q_0) \big) = \\
     &=&\lim_{t\searrow 0} \frac{1}{t} \, \EEE \big( g(Q_{0}) f(Q_{-t})
     - g(Q_{-t}) f(Q_{-t}) \big) \: +\\
     &+& \lim_{t\searrow 0} \frac{1}{t}\,\EEE \big( g(Q_{-t}) f(Q_{-t})
     - g(Q_{0}) f(Q_{0}) \big) = \\
     &=& (f,Lg) -\int\limits_{q\in\conf} g(q) \, f(q) 
\,\dot{\measure}(dq)
     \stackrel{\eqref{generatorduality}}{=} (Lg,f)-(L(gf),\1) =
     (g,L^\dag f) - (fg,L^\dag \1)\,,
   \end{eqnarray*}
   which amounts to \eqref{LT}.}
\begin{equation}\label{LT}
   \ov{L} f = L^\dag f - (L^\dag\1) f
\end{equation}
where $^\dag$ denotes the adjoint operator on $L^2(\conf,\measure)$,
with $\measure$ given by \eqref{mis}.  Since $\ov{L}$ is linear in
$L$, condition (\ref{LTinv}) is preserved when adding (forward or
backward) generators; it is also preserved under ``second
quantization.''  For a pure jump process, (\ref{LTinv}) boils down to
\begin{equation}\label{jumprevers}
   \sigma^{\Psi}(dq|q') \, \sp{\Psi}{\pov(dq') |\Psi} = \sigma^{T \Psi}
   (dq'|q) \, \sp{\Psi}{\pov(dq) |\Psi}\,,
\end{equation}
which is satisfied for the minimal jump rates, by inspection of
(\ref{tranrates}).  The minimal free generator \eqref{LH} changes sign
when replacing $\Psi$ by $T\Psi$, which means the velocity changes
sign, as it should under time reversal (see Section
\ref{sec:freeflow}).

Invariance under Galilean boosts is a more involved story, and as it
is not considered as fundamental in physics anyway, we omit it here.
Lorentz boosts are even trickier, since for more than just one
particle, they even fail to map (simultaneous) configurations into
(simultaneous) configurations.  As a result, the problem of Lorentz
invariance belongs in an altogether different league, which shall not
be entered here.

\subsection{On the Notion of Reversibility}

   It may appear, and it is in fact a widespread belief, that
   stochasticity is incompatible with time reversibility.  We naturally
   view the past as fixed, and the future, in a stochastic theory, as
   free, determined only by innovations. Even Bell expressed such a
   belief \cite[p.~177]{Bellbook}. However, from the proper perspective
   the conflict disappears, and this perspective is to consider the
   path space (of the universe) and the probability measure thereon. If
   $t\mapsto Q_t$ is a history of a universe governed by a Bell-type 
QFT, then
   its time reverse, $t\mapsto Q_{-t}$, is again a possible path of this
   Bell-type QFT, though corresponding to a different initial state
   vector $T\Psi$ instead of $\Psi$, with $T$ the time reversal
   operator as discussed in Section \ref{sec:symm}. More than this, the
   distribution of the reversed path $t\mapsto Q_{-t}$ coincides with the
   probability measure on path space arising from $T\Psi$.\footnote{We
   can be more precise about the meaning of the measure on path space:
   as  in Bohmian mechanics \cite{DGZ}, its role ``is precisely to
   permit definition of the word `typical'.'' \cite[p.~129]{Bellbook}
   Consequently, the meaning of the reversibility property of the
   measures we just mentioned is that the time reverse of a history
   that is typical with respect to $\Psi$, is typical with respect to
   $T\Psi$.}

   It may also be helpful to think of how the situation appears when
   viewed from outside space-time: then the path $Q_t$ corresponds to
   the decoration of space-time with a pattern of world lines, and this
   pattern is random with respect to a probability measure on what
   corresponds to path space, namely the space of all possible
   decorations of space-time. Then time reversal is a mere reflection,
   and for a theory to be time reversible means the same as being
   invariant under this reflection: that we could have had as well the
   reflected probability measure, provided we had started with $T\Psi$
   instead of $\Psi$.

   To sum up, we would like to convey that the sense of reversibility
   for Markov processes indeed matches the sense of reversibility that
   one should expect from a physical theory.

\subsection{Heisenberg Picture}

   In (\ref{mis}), we have applied the Schr\"odinger picture, according
   to which the state vector evolves while the operators remain
   fixed. Eq.~(\ref{mis}) and the reasoning following it can as well be
   translated to the Heisenberg picture where the state vector $\Psi$
   is regarded as fixed and the operators $\pov_t(\,\cdot\,)$ as
   evolving. Thus, we could equivalently write
   \[
     \measure_t(dq) = \sp{\Psi}{\pov_t(dq)| \Psi}
   \]
   instead of (\ref{mis}). Similarly, $H_0$ and $H_\inter$ become
   time-dependent while their sum is constant.  We often use an
   ambiguous notation like $\sp{\Psi}{\pov(dq)|\Psi}$ and formula
   \eqref{tranrates} since the formulas are equally valid in both
   pictures (and, for that matter, in the interaction picture).

   Like the jump rate formula \eqref{tranrates}, the formula \eqref{LH}
   for the minimal free generator is equally valid in the Heisenberg
   picture.

   We further remark that in the Heisenberg picture, the following nice
   equation holds for a pure jump process with minimal rates when
   $\pov$ is a PVM:
   \begin{equation}\label{twotimes}
     \prob\{Q_{t+dt} \in dq, Q_{t} \in dq'\} = \sp{\Psi} {\{\pov_{t+dt}
     (dq), \pov_{t}(dq') \} | \Psi}^+
   \end{equation}
   for $dq \cap dq' = \emptyset$, where $\{ \;,\, \}$ on the right hand
   side means the anti-commutator.  The similarity to the one-time
   distribution formula
   \[
     \prob\{Q_t \in dq\} = \sp{\Psi}{\pov_t(dq) |\Psi}
   \]
   is striking.  Specifying the two-time distribution for infinitesimal
   time differences is a way of characterizing a Markov process,
   equivalent to specifying the (forward or backward) generator and the
   one-time distribution.  Thus, for a PVM $\pov$ \eqref{twotimes}
   provides another formula for the minimal jump rates
   \eqref{tranrates}. A similar formula for the process generated by
   the minimal free generator \eqref{LH} is $\EEE \big(g(Q_t)
   f(Q_{t+dt}) \big) = \frac12 \sp{\Psi} {\{\hat{g}_t, \hat{f}_{t+dt}
   \} | \Psi}$.

\subsection{Examples of Process Additivity}
\label{sec:known}

Among different versions of Bohmian mechanics we find numerous
examples of process additivity (and, remarkably, no example
\emph{violating} it):
\begin{itemize}
\item The Hamiltonian for $n$ noninteracting particles is the sum of
   the Hamiltonians for the individual particles, and it is easy to see
   that the vector field defining Bohmian mechanics for the
   $n$-particle system is the sum of the vectors fields (each regarded
   as vectors fields on $\RRR^{3n}$) for the particles. As already
   mentioned, sums of generators for deterministic processes amount to
   sums of the defining vector fields.

   Moreover, the vector field for each particle is essentially the
   Bohmian one-particle law. To point out that this is a nontrivial
   fact, we mention that this is not so for the alternative velocity
   formula (10.2) in \cite{Deotto} considered by Deotto and Ghirardi,
   for which the velocity of the $i$-th particle differs from the
   one-particle law. So Bohmian mechanics of $n$ particles can be
   viewed as built from $n$ copies of the one-particle version, in fact
   by the ``second quantization'' algorithm of Section \ref{sec:Gamma}.

\item The vector field of Bohmian mechanics for a single spinless
   particle may also be seen as arising in this way. If a Hamiltonian
   $H=-X^2$ is the negative square of an (incompressible) vector field
   (regarded as a first-order differential operator) $X=a(\vx) \!\cdot\!
   \nabla$ on $\RRR^3$ (with $\nabla \!\cdot\! a=0$ ensuring formal
   self-adjointness of the square), then the simplest equivariant
   process associated with $H$ is given by the velocity vector field
   \[
      v= \frac{2}{\hbar} \, \Im \,\frac{a\cdot \nabla \Psi}{\Psi}\, a\, .
   \]
   The corresponding backward generator is $L = \frac{2}{\hbar} \, \Im
   \, (\frac{X\Psi}{\Psi}) X$.  Now $-\frac{\hbar^2}{2}\Laplace =
   -\sum_{\alpha}{X_{\alpha}}^2$ is the sum of 3 negative squares of
   vector fields $X_{\alpha} = \frac{\hbar} {\sqrt{2}} \partial /
   \partial x^\alpha$ corresponding to the individual degrees of
   freedom. The associated Bohm velocity is the sum of the velocities
   corresponding to the squares. So Bohmian mechanics in three
   dimensions can be viewed as built of 3 copies of the one-dimensional
   version. To point out that this is a nontrivial fact, we mention
   that this is not true, e.g., of the velocity formulas (10.1) and
   (10.2) in \cite{Deotto}, which do not make sense in dimensions other
   than 3.

\item If we add an interaction potential $V$ to $-\frac{\hbar^2}{2}
   \Laplace$, the Bohm velocity is the appropriate sum, since the
   operator $V$ is associated with the trivial motion $v=0$.

\item We may also include an external vector potential $\vA(\vx,t)$ in
   the Schr\"odinger equation, that is, replace $- \frac{\hbar^2}{2}
   \Laplace = - \frac{\hbar^2}{2} \nabla^2$ by $- \frac{\hbar^2}{2}
   \big( \nabla + \I \frac{e}{\hbar} \vA(\vx,t) \big)^2 = -
   \frac{\hbar^2}{2} \Laplace - \frac{\hbar^2}{2} (\I \frac{e}{\hbar}
   \nabla\cdot\vA + \I \frac{e}{\hbar} \vA \cdot \nabla) +
   \frac{e^2}{2} \vA^2$. The sum of the associated velocities, namely
   \[
     \hbar \, \Im \, \frac{\Psi^* \nabla \Psi}{\Psi^* \, \Psi} + e\vA +
     0
   \]
   equals the velocity one obtains directly, $\hbar \, \Im \, \Psi^*
   (\nabla +\I \frac{e}{\hbar} \vA)\Psi/ \Psi^* \, \Psi$.

\item In the Bohm--Dirac theory (\ref{BohmDirac}), however, one can
   include an external gauge connection $A_\mu(\vx,t)$ in the Dirac
   equation without changing the velocity formula. That conforms with
   process additivity because the operator $(\gamma^0)^{-1} \gamma^\mu
   A_\mu = A_0+\boldsymbol{\alpha}\cdot\vA$ is associated (termwise)
   with $v=0$.

\item In the Dirac Hamiltonian $H = -\I c \hbar \valpha \cdot \nabla +
   \beta mc^2$, the first term corresponds to the Bohm--Dirac velocity
   (\ref{BohmDirac}), whereas the second term corresponds to $v=0$; as
   a consequence, the Bohm--Dirac velocity does not depend on the mass.
   Moreover, the three components of the Bohm--Dirac velocity are each
   equivariant with respect to the corresponding derivative term in
   $H$.
\end{itemize}

In addition, we point out cases of process additivity in the ``second
quantization'' algorithm and minimal jump processes.


The ``second quantized'' generator $\Gamma \generator^{(1)}$ as
constructed in Section \ref{sec:Gamma} provides an example of
process additivity (or may be viewed as an application of process
additivity):
\[
   \generator_{H_0, \Psi}= \sum_{n=0}^{\infty}
   \generator_{H_0^{(n)}, \Psi^{(n)}} \,,
\]
where the generators in the sum correspond to motions in the
respective different sectors of $\conf$.

Suppose we regard the particles as ordered, $Q = (\vQ_1, \ldots,
\vQ_N)$. Then another case of process additivity becomes visible:
\[
   H_0^{(N)} = \sum_{i=1}^N h_i
\]
where $h_i$ is the one-particle Hamiltonian acting on the $i$-th
particle. Correspondingly,
\[
   \generator_{H_0^{(N)}} = \sum_{i=1}^N \generator_i
\]
where $\generator_i$ is equivariant with respect to $h_i$. This
applies not only to Bohmian mechanics (as described earlier in this
section), but generally to the ``second quantization'' procedure as
described in Section \ref{sec:Gamma}.  We also note that the ``second
quantization'' algorithm presented in Section \ref{sec:Gamma}
preserves process additivity in the sense that
$\Gamma(\generator_1^{(1)} + \generator_2^{(1)}) =
\Gamma(\generator_1^{(1)}) + \Gamma(\generator_2^{(1)})$ while
$\Gamma(H_1^{(1)} + H_2^{(1)}) = \Gamma(H_1^{(1)}) +
\Gamma(H_2^{(1)})$.

\label{sec:miniadd}

We now turn to process additivity among minimal jump processes.

A jump process generated by a sum need not be a minimal jump process
even when its constituents are.  But under certain conditions it is.
Two such cases are the ``direct sum'' and ``tensor product'' processes
constructed in Sections \ref{sec:directsum} and
\ref{sec:tensorproduct}: $\Hilbert = \bigoplus_n \Hilbert^{(n)}$ with
$\conf = \bigcup_n \conf^{(n)}$ and $H =\bigoplus_n H^{(n)}$, and
$\Hilbert = \Hilbert^{[1]} \otimes \cdots \otimes \Hilbert^{[N]}$ with
$\conf = \conf^{[1]} \times \cdots \times \conf^{[N]}$ and $H = \sum_i
\1 \otimes \cdots \otimes H^{[i]} \otimes \cdots \otimes \1$, with
$\generator = \sum \generator_i$ where $\generator_i$ acts
nontrivially, in an obvious sense, only on $\conf^{(i)}$ or on
$\conf^{[i]}$.  These are special cases of the general fact that
minimality is compatible with additivity whenever the addends of the
Hamiltonian correspond to \emph{different sorts} of jumps. That can be
most easily understood in the case of a PVM corresponding to an
orthonormal basis $\{|q\rangle : q \in \conf\}$ of $\Hilbert$: suppose
$H=H_1 + H_2$ and for every pair $q,q'$ either $\sp{q}{H_1|q'} =0$ or
$\sp{q}{H_2 |q'} =0$. Then $\sigma = \sigma_1 + \sigma_2$.  The
corresponding condition in the POVM context is that the kernels of
$H_1$ and $H_2$ have disjoint supports. When $H$ is naturally given as
a sum this condition would be expected to be satisfied.

Finally, we remark that the minimal free generator $\generator =
\generator^H$ as defined in (\ref{genH}) is additive in $H$.

\label{sec:mini3}

\subsection{Second Quantization of a Minimal Jump Process}

   We note that the ``second quantization'' of a minimal jump
   process associated with a PVM $\pov^{(1)}$, as described in Section
   \ref{sec:Gamma}, is the minimal jump process associated with the
   second-quantized Hamiltonian; this is a consequence of the
   observation that $\generator_i$ generates the minimal jump process
   for $H_i$ in this case. This fact is probably physically irrelevant
   but it is mathematically nice.

\subsection{Global Existence Question}

The rates $\sigma_t$ and velocities $v_t$, together with $\measure_t$,
define the process $Q_t$ associated with $H,\pov$, and $\Psi$, which
can be constructed along the lines of Section
\ref{sec:revjump}. However, the rigorous existence of this process,
like the global existence of solutions for an ordinary differential
equation, is no trivial matter.  See Section~4.3 of \cite{crea2A} for
a discussion of what must be controlled in order to establish the
global existence of the process, and \cite{crex1} for an example of
such a global existence proof.

\subsection{POVM Versus PVM}

As we have already remarked in footnote \ref{ft:Naimark}, every POVM
$\pov$ is related to a PVM $\pov_\ext$, the Naimark extension, on a
larger Hilbert space $\Hilbert_\ext$ according to $\pov(\,\cdot\,) =
P_+ \pov_\ext(\,\cdot\,) I$ with $P_+$ the projection $\Hilbert_\ext
\to \Hilbert$ and $I$ the inclusion $\Hilbert \to \Hilbert_\ext$. This
fact allows a second perspective on $\pov$, and sometimes creates a
certain ambiguity as to which process is the suitable one for a
Bell-type QFT, as follows.  At several places in this paper, we have
described considerations leading to and methods for defining Markov
processes, in particular minimal jump rates \eqref{tranrates} and the
minimal free generator \eqref{LH}; these considerations and methods
could be applied using either $\Hilbert_\ext$ and $\pov_\ext$ or
$\Hilbert$ and $\pov$. One would insist that the state vector $\Psi$
must lie in $\Hilbert$, the space of physical states, but even then
one might arrive at different processes starting from $\pov$ or
$\pov_\ext$. To obtain a process from $\pov_\ext$ requires, of course,
that we have a Hamiltonian on $\Hilbert_\ext$, while $H$ is defined on
$\Hilbert$; such a Hamiltonian, however, can easily be constructed
from $H$ by setting $H_\ext = I H P_+$.

In some cases, the Naimark extension does not lead to an ambiguity.
This is the case for the jump rate formula \eqref{tranrates}, since
for $\Psi \in \Hilbert$, $\sp{\Psi}{\pov_\ext(dq)| \Psi} =
\sp{\Psi}{\pov(dq) |\Psi}$ and $\sp{\Psi}{\pov_\ext(dq) H_\ext
\pov_\ext(dq')| \Psi} = \sp{\Psi}{\pov(dq) H \pov(dq') |\Psi}$. This
fact suggests that,  generally, the minimal process arising from
$H_\ext$ and $\pov_\ext$ is the same as the one arising from $H$ and
$\pov$.

The situation is different, however, when $H$ is defined on
$\Hilbert_\ext$ to begin with, and different from $H_\ext$. This is
the case with the free Dirac operator $h_0$, defined as a differential
operator on $L^2(\RRR^3,\CCC^4)$, which differs from $P_+h_0P_+$.
When we obtained in Section \ref{sec:free2} the Bohm--Dirac motion
\eqref{BohmDirac} from the formula \eqref{LH} for the minimal free
generator, we used $h_0$ and $\pov_\ext$. In contrast, the restriction
of $h_0$ to the positive energy subspace, or equivalently $P_+ h_0
P_+$, possesses a kernel; more precisely, it is a convolution operator
$S_+ \star (h_0 S_+) \star$ in the notation of Section
\ref{sec:positron}, and thus corresponds to jumps. The associated
minimal process on $\RRR^3$ presumably makes infinitely many jumps in
every finite time interval, similar to the example of \cite{crea2A}, 
Section
3.5.

Thus, there are two processes to choose between, the Bohm--Dirac
motion and the minimal process for $P_+ h_0 P_+$. Both are
equivariant, and thus it is arguably impossible to decide empirically
which one is right.  In our example theory in Section
\ref{sec:positron}, we chose the simpler, deterministic one. But we
leave to future work the discussion of which is more likely relevant
to physics, and why.

\subsection{The Role of Field Operators}\label{sec:fields}

The Bell-type QFTs with which we have been concerned in this paper are
models describing the behaviour of \emph{particles} moving in physical
3-space, not of fields on 3-space. We have been concerned here mainly
with a particle ontology, not a field ontology. This focus may be
surprising at first: almost by definition, it would seem that QFT
deals with fields, and not with particles.  Consider only the
occurrence (and prominence) of field operators in QFT!

But there is less to this than might be expected.  The field operators
do not function as observables in QFT. It is far from clear how to
actually ``observe'' them, and even if this could somehow, in some
sense, be done, it is important to bear in mind that the standard
predictions of QFT are grounded in the particle representation, not
the field representation: Experiments in high energy physics are
scattering experiments, in which what is observed is the asymptotic
motion of the outgoing particles.  Moreover, for Fermi fields---the
matter fields---the field as a whole (at a given time) could not
possibly be observable, since Fermi fields anti-commute, rather than
commute, at space-like separation. One should be careful here
not to be taken in by the attitude widespread in quantum theory
of intuitively regarding the operators as ``quantities,'' as if they
represented something out there in reality; see \cite{naive} for a
critique of this attitude.

So let us focus on the role of the field operators in QFT.  This seems
to be to relate abstract Hilbert space to space-time: the field
operators are attached to space-time points, unlike the quantum states
$\Psi$, which are usually regarded not as functions but as abstract
vectors. In orthodox quantum field theory the field operators are an
effective device for the specification of Hamiltonians having good
space-time properties. For our purposes here, what is critical is the
connection between field operators and POVMs.

Throughout this paper, the connection between Hilbert space and the
particle positions in physical space has been made through the POVM
$\pov$, and through it alone.  We now wish to emphasize that the field
operators are closely related to $\pov$, and indeed that field
operators are just what is needed for efficiently defining a POVM
$\pov$ on $\Gamma(\RRR^3)$.

This connection is made through number operators $N(R)$, $R \subseteq
\RRR^3$.  These define a \emph{number-operator-valued measure} (NOVM)
$N(\,\cdot\,)$ on $\RRR^3$, an ``unnormalized POVM'' ($N(\RRR^3)$ is
usually not the identity operator and $N(R)$ is usually an unbounded
positive operator) for which the values $N(R)$ commute and are number
operators: $\mathrm{spectrum}(N(R)) \subseteq \{0,1,2,3,\ldots\}$.
(The basic difference, then, between a NOVM and a PVM is that the
spectrum of the positive operators is $\{0,1,2,3,\ldots\}$ rather than
just $\{0,1\}$.)

There is an obvious one-to-one relation between NOVMs $N(\,\cdot\,)$
on $\RRR^3$ and PVMs $\pov$ on $\Gamma(\RRR^3)$, given by
\begin{equation}\label{Npov}
   N(R) = \int\limits_{q\in\Gamma(\RRR^3)} n_R(q) \, \pov(dq)
\end{equation}
where $n_R(q) = \#(q \cap R)$ is the number function on
$\Gamma(\RRR^3)$ for the region $R$. Since \eqref{Npov} is the
spectral decomposition of the commuting family $N(R)$, this
correspondence is one-to-one. (Note that the joint spectrum of the
commuting family $N(R)$ is the set of nonnegative-integer-valued
measures $n_R$ on $\RRR^3$, one of the definitions of $\Gamma(\RRR^3)$
given in Section \ref{sec:free}.)

The moral is that a NOVM on $\RRR^3$ is just a different way of
speaking about a PVM $\pov$ on $\conf = \Gamma(\RRR^3)$.  All other
POVMs arise from PVMs by restriction to a subspace (Naimark's theorem
\cite{Davies}).  An easy way to obtain a NOVM $N$ starts with setting
\begin{equation}\label{Ndef}
   N(R) = \int_R \phi^*(\vx) \, \phi(\vx) \, d^3\vx
\end{equation}
for suitable operators $\phi(\vx)$.  An easy way to ensure that the 
$N(R)$
commute is to require that the operators $\phi(\vx)$ commute or
anti-commute
with each other and the adjoints $\phi^*(\vx')$ for $\vx'\neq \vx$.
An easy way to ensure that the $N(R)$ have nonnegative integer 
eigenvalues
is to require that
\begin{equation}\label{ccr}
   [\phi(\vx),\phi^*(\vx')]_\pm = \delta(\vx-\vx')\,,
\end{equation}
where $[ \;,\,]_\pm$ is the (anti\nobreakdash-)commutator, and that
there is a cyclic vacuum state $|0\rangle \in \Hilbert$ for which
$\phi(\vx)|0 \rangle =0$.  The relations \eqref{ccr} are of course
just the usual canonical (anti\nobreakdash-)commutation relations that
field operators are required to satisfy.

Moreover, in gauge theories the connection between matter field $\phi$
and the NOVM is perhaps even more compelling. Consider a gauge theory
with internal state space $V$, equipped with the inner product
$\scalar{\,\cdot\,}{\,\cdot\,}$. Then, given $\vx \in \RRR^3$, the
matter field $\phi(\vx)$ should formally be regarded as a linear
functional $V \to \mathcal{O}(\Hilbert)$, $\xi \mapsto \phi_\xi(\vx)$,
from the internal state space to operators on $\Hilbert$, with
$\phi^*_{\xi^*}(\vx) = (\phi_\xi(\vx))^*$ a linear function $V^* \to
\mathcal{O}(\Hilbert)$ on the dual of $V$. \eqref{ccr} then becomes
$[\phi_\xi(\vx), \phi_{\eta^*}^*(\vx')] = \delta(\vx-\vx') \,
\scalar{\eta}{\xi}$. Thus the simplest gauge-invariant object
associated with $\phi$ is the NOVM \eqref{Ndef}, with the integrand
understood as the contraction of the tensor $V \times V^* \to
\mathcal{O}(\Hilbert)$, $(\xi,\eta) \mapsto \phi_\eta^*(\vx) \,
\phi_\xi(\vx)$.

Hence, not only does the notion of particle not conflict with the
prominence of field operators (see Sections \ref{sec:crea1} and
\ref{sec:positron} for explicit examples), but field operators have a
natural place in a theory whose ultimate goal it is to govern the
motion of particles.  One of their important roles is to define the
POVM $\pov$ that relates Hilbert space to configuration space.
Quantum theory of fields or quantum theory of particles? A theory of
particle motion exploiting field operators!

\section{Conclusions}

The essential point of this paper is that there is a direct and
natural way of understanding QFT as a theory about moving particles,
an idea pioneered, in the realm of nonrelativistic quantum mechanics,
by de Broglie and Bohm.  We leave open, however, three considerable
gaps: the question of the process associated with the Klein--Gordon
operator, the problem of removing cut-offs, and the issue of Lorentz
invariance.

\bigskip

\noindent \textbf{Acknowledgements. }We thank James Taylor of Rutgers
University and Stefan Teufel of Technische Universit\"at M\"unchen for
helpful discussions.  R.T.\ gratefully acknowledges support by the
German National Science Foundation (DFG).  N.Z.\ gratefully
acknowledges support by INFN and DFG.  Finally, we appreciate the
hospitality that some of us have enjoyed, on more than one occasion,
at the Mathematisches Institut of Ludwig-Maximilians-Universit\"at
M\"unchen, at the Dipartimento di Fisica of Universit\`a di Genova,
and at the Mathematics Department of Rutgers University.


\begin{thebibliography}{28.}

\bibitem{ali} Ali, S.T., Emch, G.G.: Fuzzy Observables in Quantum
   Mechanics, J.\ Math.\ Phys.\ \textbf{15}, 176-182 (1974)


\bibitem{BD} Bacciagaluppi, G., Dickson, M.: Dynamics for modal
   interpretations, Found.\ Phys.\ \textbf{29},
   1165-1201 (1999), and quant-ph/9711048

\bibitem{BellBeables} Bell, J.S.: Beables for quantum field
   theory, Phys.\ Rep.\ \textbf{137}, 49-54 (1986). Reprinted in
   \cite{Bellbook}, p.~173.

\bibitem{Belldensity} Bell, J.S.: De Broglie--Bohm, delayed-choice
   double-slit experiment, and density matrix, Int.\ J.\ Quant.\
   Chem.\ \textbf{14}, 155-159 (1980). Reprinted in \cite{Bellbook},
   p.~111.

\bibitem{Bellbook} Bell, J.S.: \textit{Speakable and unspeakable in
   quantum mechanics}.  Cambridge: Cambridge University Press (1987)


\bibitem{Bohm52} Bohm, D.: A Suggested Interpretation of the Quantum
   Theory in Terms of ``Hidden'' Variables, I, Phys.\ Rev.\
   \textbf{85}, 166-179 (1952).  Bohm, D.: A Suggested Interpretation
   of the Quantum Theory in Terms of ``Hidden'' Variables, II, Phys.\
   Rev.\ \textbf{85}, 180-193 (1952)

\bibitem{BH} Bohm, D., Hiley, B.J.: \textit{The Undivided Universe: An
   Ontological Interpretation of Quantum Theory}. London: Routledge,
   Chapman and Hall (1993)

\bibitem{Breiman} Breiman, L.: \textit{Probability}. Reading:
   Addison-Wesley (1968)

\bibitem{naive} Daumer, M., D\"urr, D., Goldstein, S., Zangh{\`\i},
   N.: Naive Realism about Operators, Erkenntnis \textbf{45},
   379-397 (1996), and quant-ph/9601013

\bibitem{Davidson} Davidson, M.: A generalization of the
   F\'enyes--Nelson stochastic model of quantum mechanics, Lett.\
   Math.\ Phys.\ \textbf{3}, 271-277 (1979)

\bibitem{Davies} Davies, E.B.: \textit{Quantum Theory of Open Systems}.
   London, New York, San Francisco: Academic Press (1976)

\bibitem{Deotto} Deotto, E., Ghirardi, G.C.: Bohmian mechanics
   revisited, Found.\ Phys.\ \textbf{28}, 1-30 (1998), and
   quant-ph/9704021



\bibitem{identical} D\"urr, D., Goldstein, S., Taylor, J., Tumulka,
   R., Zangh{\`\i}, N.: Bosons, Fermions, and the Topology of
   Configuration Space, in preparation.

\bibitem{crea1} D\"urr, D., Goldstein, S., Tumulka, R., Zangh{\`\i},
   N.: Trajectories and Particle Creation and Annihilation in Quantum
   Field Theory, J.\ Phys.\ A: Math.\ Gen.\ \textbf{36}, 4143-4149
   (2003), and quant-ph/0208072

\bibitem{crea2A} D\"urr, D., Goldstein, S., Tumulka, R., Zangh{\`\i},
   N.: Quantum Hamiltonians and Stochastic Jumps, quant-ph/0303056

\bibitem{crea3} D\"urr, D., Goldstein, S., Tumulka, R., Zangh{\`\i},
   N.: Quantum Theory and Minimal Processes, in preparation.

\bibitem{crea4} D\"urr, D., Goldstein, S., Tumulka, R., Zangh{\`\i},
   N.: QED With Particles, in preparation.

\bibitem{klein2} D\"urr, D., Goldstein, S., Tumulka, R., Zangh{\`\i},
   N.: Trajectories From Klein--Gordon Functions, in preparation.

\bibitem{DGZ} D\"urr, D., Goldstein, S., Zangh{\`\i}, N.: Quantum
   equilibrium and the origin of absolute uncertainty, J.\ Statist.\
   Phys.\ \textbf{67}, 843-907 (1992)

\bibitem{crex1} Georgii, H.-O., Tumulka, R.: Global Existence of Bell's
   Time-Inhomogeneous Jump Process for Lattice Quantum Field Theory, to
   appear in \textit{Markov Processes Rel.\ Fields} (2004), and
   math.PR/0312294

\bibitem{Stanford} Goldstein, S.: Bohmian Mechanics (2001), in:
   \textit{Stanford Encyclopedia of Philosophy (Winter 2002 Edition)},
   E.N.~Zalta (ed.), \newline
   http://plato.stanford.edu/archives/win2002/entries/qm-bohm/

\bibitem{Guerra} Guerra, F., Marra, R.: Discrete stochastic
   variational principles and quantum mechanics, Phys.~Rev.~D
   \textbf{29}, 1647-1655 (1984)


\bibitem{Haag} Haag, R.: \textit{Local Quantum Physics: Fields,
   Particles, Algebras}.  Berlin: Springer-Verlag (1992)



\bibitem{Jaekel} Jaekel, M.T., Pignon, D.: Stochastic Processes of a
   Quantum State, Int.\ J.\ Theor.\ Phys.\ \textbf{24}, 557-569
   (1985)

\bibitem{kraus} Kraus, K.: Position Observables of the Photon,
   p.~293-320 in W.C.~Price and S.S.~Chissick (eds.), \textit{The
   Uncertainty Principle and Foundations of Quantum Mechanics}. New
   York: Wiley (1977)



\bibitem{stochmech} Nelson, E.: \textit{Quantum Fluctuations}.
   Princeton: Princeton University Press (1985)

\bibitem{NewtonWigner} Newton, T.D., Wigner, E.P.: Localized States
   for Elementary Systems, Rev.\ Mod.\ Phys.\ \textbf{21}, 400-406
   (1949)

\bibitem{photon} Norsen, T., Tumulka, R.: A Model of Photon
   Trajectories, in preparation.


\bibitem{RS} Reed, M., Simon, B.: \textit{Methods of Modern
   Mathematical Physics. I: Functional Analysis}. New York and London:
   Academic Press (1972)

\bibitem{Roy} Roy, S.M., Singh, V.: Generalized beable quantum
   field theory, Phys.\ Lett.\ B \textbf{234}, 117-120 (1990)

\bibitem{Ruijsenaars} Ruijsenaars, S.N.M.: Charged Particles in
   External Fields. I. Classical Theory, J.\ Math.\ Phys.\
   \textbf{18} No 4, 720-737 (1977), and Charged Particles in
   External Fields. II. The Quantized Dirac and Klein--Gordon
   Theories, Commun.\ Math.\ Phys.\ \textbf{52}, 267-294 (1977)

\bibitem{Schweber} Schweber, S.S.: \textit{An Introduction to
   Relativistic Quantum Field Theory}. New York: Harper and Row (1961)

\bibitem{Sudbery} Sudbery, A.: Objective interpretations of quantum
   mechanics and the possibility of a deterministic limit, J.\ Phys.\
   A: Math.\ Gen.\ \textbf{20}, 1743-1750 (1987)


\bibitem{Vink} Vink, J.C.: Quantum mechanics in terms of discrete
   beables, Phys.\ Rev.\ A \textbf{48}, 1808-1818 (1993)



\end{thebibliography}
\end{document}